\documentclass[nofootinbib,prb,reprint,twocolumn,amsmath,amssymb]{revtex4-1}

\newcommand{\parder}[2]{\frac{\partial #1}{\partial #2}}

\newcommand{\shp}[2]{N^{#1}_{#2}(\bx)}

\newcommand{\del}{\nabla}

\newcommand{\bx}{\boldsymbol{\textbf{x}}}

\newcommand{\bR}{\boldsymbol{\textbf{R}}}

\newcommand{\intomega}{\int_{\Omega}}
\newcommand{\intomegai}{\int_{\Omega_{I}}}

\newcommand{\bk}{\boldsymbol{\textbf{k}}}

\newcommand{\ualbk}{u_{\alpha,\mathbf{k}}}

\newcommand{\epsalbkBar}{\bar{\epsilon}_{\alpha,\mathbf{k}}}
\usepackage{subfiles}
\usepackage{graphicx}
\usepackage{graphics}
\usepackage{multirow}
\usepackage{fancyhdr, ifpdf}
\usepackage{amsxtra}
\usepackage{stmaryrd}
\usepackage{mathrsfs}
\usepackage{mathtools}
\usepackage{psfrag}
\usepackage{threeparttable}
\usepackage{subfigure}
\usepackage{epsfig}
\usepackage{rotating}
\usepackage{setspace}
\usepackage{float}
\usepackage{amsfonts}
\usepackage{amsbsy}
\usepackage{amscd}
\usepackage{amsthm}
\usepackage{color}
\usepackage{tabularx}
\usepackage{sidecap}
\usepackage{dcolumn}
\usepackage{footnote}
\usepackage{algorithm}
\usepackage{algorithmic}
\usepackage{enumitem}
\usepackage[titletoc,title]{appendix}

\usepackage{comment}
\usepackage{soul}

\definecolor{hellgruen}{rgb}{0.2,0.7,0.2}

\newcolumntype{M}[1]{>{\centering\arraybackslash}m{#1}}
\newcolumntype{N}{@{}m{0pt}@{}}

\graphicspath{ {./images/} }
\newcommand{\epsder}[1]{\left. \frac{d}{d\varepsilon} #1 \right \vert_{\varepsilon=0}}

\usepackage{esint} 

\newcommand{\eps}[1]{#1^{\varepsilon}}

\newcommand{\hbarr}[1]{\bar{#1}^{h}}
\newcommand{\evalepszero}{\bigg|_{\varepsilon=0}}
\newcommand{\ddeps}{\frac{d}{d\varepsilon}}

\newcommand{\shpR}[3]{N^{#1}_{#2}(#3)}

\newcommand{\blkcomment}[1]{}

\newcommand{\shpFcnTemplate}[3]{N^{#1}_{#2}(#3)}
\newcommand{\shpFuncClass}[1]{\shpFcnTemplate{C}{#1}{\bx}}
\newcommand{\shpFuncClassEps}[1]{\shpFcnTemplate{C,\varepsilon}{#1}{\bx^\varepsilon}}
\newcommand{\shpFuncEnrichment}[3]{\shpFcnTemplate{E,#2}{#1}{\bx - \bR_{#3}}}
\newcommand{\shpFuncEnrichmentEps}[3]{\shpFcnTemplate{E,#2}{#1}{\bx^\varepsilon - \bR_{#3}^\varepsilon}}
\newcommand{\shpFuncPhiEnrichment}[1]{\shpFuncEnrichment{#1}{\phi}{#1}}
\newcommand{\shpFuncPhiEnrichmentEps}[1]{\shpFuncEnrichmentEps{#1}{\phi}{#1}}
\newcommand{\shpFuncUEnrichment}[2]{\shpFuncEnrichment{#1,#2}{u_{\bk}}{#2}}
\newcommand{\shpFuncUEnrichmentEps}[2]{\shpFuncEnrichmentEps{#1,#2}{u_{\bk}}{#2}}

\newcommand{\coeffUClassical}[2]{u_{#1,\bk,#2}^C}

\newcommand{\coeffUClassicalBar}[2]{\bar{u}_{#1,\bk,#2}^C}
\newcommand{\coeffUClassicalEpsBar}[2]{\bar{u}_{#1,\bk,#2}^{C,\varepsilon}}

\newcommand{\coeffUEnrichment}[3]{u_{#1,\bk,#2,#3}^E}

\newcommand{\coeffUEnrichmentBar}[3]{\bar{u}_{#1,\bk,#2,#3}^E}
\newcommand{\coeffUEnrichmentEpsBar}[3]{\bar{u}_{#1,\bk,#2,#3}^{E,\varepsilon}}

\newcommand{\coeffPhiClassical}[1]{\phi_{#1}^C}

\newcommand{\coeffPhiClassicalBar}[1]{\bar{\phi}_{#1}^C}
\newcommand{\coeffPhiClassicalEpsBar}[1]{\bar{\phi}_{#1}^{C,\varepsilon}}

\newcommand{\coeffPhiEnrichment}[1]{\phi_{#1}^E}

\newcommand{\coeffPhiEnrichmentBar}[1]{\bar{\phi}_{#1}^E}
\newcommand{\coeffPhiEnrichmentEpsBar}[1]{\bar{\phi}_{#1}^{E,\varepsilon}}

\newcommand{\wfnGrpAtKEpsBar}{\bar{\mathbf{u}}^\varepsilon_{\bk}} 
\newcommand{\wfnDiscNumK}[2]{u_{#1,#2}^{h}} 
\newcommand{\wfnDiscNumKConj}[2]{u_{#1,#2}^{h,*}} 
\newcommand{\wfnDiscNumKBar}[2]{\bar{u}_{#1,#2}^{h}} 
\newcommand{\wfnDiscNumKConjBar}[2]{\bar{u}_{#1,#2}^{h,*}} 
\newcommand{\wfnTildeDiscNumKEvalAt}[6]{u_{#1,#2}^{h}\Big(#3,#4,#5,#6\Big)} 
\newcommand{\wfnTildeDiscNumK}[2]{u_{#1,#2}^{h}} 
\newcommand{\wfnTildeDiscNumKConj}[2]{u_{#1,#2}^{h,*}} 

\newcommand{\wfnSpecialEpsDer}{\left. \frac{d}{d\varepsilon}\widetilde{\ualbk^{h,\varepsilon}}(\bx^\varepsilon) \right \vert_{\varepsilon=0}}
\newcommand{\gradWfnSpecialEpsDer}{\left. \frac{d}{d\varepsilon}\del\widetilde{\ualbk^{h,\varepsilon}}(\bx^\varepsilon) \right \vert_{\varepsilon=0}}
\newcommand{\wfnSpecialEpsDerConj}{\left. \frac{d}{d\varepsilon}\widetilde{\ualbk^{h,\varepsilon,*}}(\bx^\varepsilon) \right \vert_{\varepsilon=0}}
\newcommand{\gradWfnSpecialEpsDerConj}{\left. \frac{d}{d\varepsilon}\del\widetilde{\ualbk^{h,\varepsilon,*}}(\bx^\varepsilon) \right \vert_{\varepsilon=0}}

\newcommand{\phiEpsBar}{\bar{\phi}^\varepsilon}
 
\newcommand{\phiTildeDiscEvalAt}[4]{\phi^{h}\Big(#1,#2,#3,#4\Big)} 
\newcommand{\phiTildeDisc}{\phi^{h}} 
\newcommand{\phiDisc}{\phi^h}

\newcommand{\phiSpecialEpsDer}{\left. \frac{d}{d\varepsilon}\widetilde{\phi^{h,\varepsilon}}(\bx^\varepsilon) \right \vert_{\varepsilon=0}}
\newcommand{\gradPhiSpecialEpsDer}{\left. \frac{d}{d\varepsilon}\del\widetilde{\phi^{h,\varepsilon}}(\bx^\varepsilon) \right \vert_{\varepsilon=0}}

\newcommand{\rhoDisc}{\rho^h}
\newcommand{\rhoDiscBar}{\bar{\rho}^h}

\newcommand{\densOpMat}{\boldsymbol{\Gamma}^{\mathbf{u}_{\bk}}}

\newcommand{\densOpMatElems}[2]{\Gamma_{#1 #2}^{\mathbf{u}_{\bk}}}
\newcommand{\densOpMatBar}{\bar{\boldsymbol{\Gamma}}^{\mathbf{u}_{\bk}}}
\newcommand{\densOpMatEpsBar}{\bar{\boldsymbol{\Gamma}}^{\mathbf{u}_{\bk},\varepsilon}}

\newcommand{\collectionOf}[1]{\{#1\}}

\newcommand{\evalAtEpsZero}{\bigg|_{\varepsilon=0}}

\newcommand{\integrateBZ}[1]{\fint_{\Omega_{\text{BZ}}}#1d\bk}
\newcommand{\sMatElem}[2]{S^{\bk}_{#1 #2}}
\newcommand{\sMatElemInv}[2]{{S^{\bk}_{#1 #2}}^{-1}}
\newcommand{\sMat}{S^{\bk}}
\newcommand{\sMatInv}{{S^{\bk}}^{-1}}

\newcommand{\funcDer}[2]{\frac{\delta#1}{\delta#2}}
\newcommand{\kronDelta}[2]{\delta_{#1#2}}
\newcommand{\fracOcc}[2]{f_{#1,#2}}

\newcommand{\matIdentity}{I}

\begin{document}
\raggedbottom
\title{Ionic forces and stress tensor in all-electron DFT calculations using enriched finite element basis}
\author{Nelson D. Rufus}
\affiliation{Department of Mechanical Engineering, University of Michigan, Ann Arbor, Michigan 48109, USA}
\author{Vikram Gavini}
\affiliation{Department of Mechanical Engineering, University of Michigan, Ann Arbor, Michigan 48109, USA}
\affiliation{Department of Materials Science \& Engineering, University of Michigan, Ann Arbor, Michigan 48109, USA}
\begin{abstract}

The enriched finite element basis---wherein the finite element basis is enriched with  atom-centered numerical functions---has recently been shown to be a computationally efficient basis for systematically convergent all-electron DFT ground-state calculations. In this work, we present the expressions to compute variationally consistent ionic forces and stress tensor for all-electron DFT calculations in the enriched finite element basis. In particular, we extend the formulation of configurational forces in [Motamarri \& Gavini (2018)]~\cite{Motamarri2018} to the enriched finite element basis and elucidate the additional contributions arising from the enrichment functions. We demonstrate the accuracy of the formulation by comparing the computed forces and stresses for various benchmark systems with  those obtained from finite-differencing the ground-state energy.  Further, we also benchmark our calculations against Gaussian basis for molecular systems and against the LAPW+lo basis for periodic systems.

\end{abstract}
\maketitle

\section{\label{sec:intro} Introduction}

Density functional theory~\cite{dftHohKohn}(DFT) has been the workhorse of electronic structure calculations for many decades, providing important qualitative and quantitative insights into many materials properties. The success of DFT is attributed to the reduced computational complexity---cubic scaling with system size---via the Kohn-Sham (KS) formulation~\cite{dftKohnSham}, which reduces the many-body Schr\"{o}dinger equation to an effective single-electron problem for ground-state properties. Thus, the electronic ground-state, for given positions of nuclei,  can be computed by solving the Kohn-Sham eigenvalue problem. However, computing the ground-state energy of the system also requires structural relaxations, which necessitates the evaluation of ionic forces, and, in the context of periodic geometries, also the stress tensor associated with the cell geometry. 

The ionic force on a nucleus is simply the negative derivative of the electronic ground-state energy with respect to the position of the nucleus. The dependence of the electronic ground-state energy on the position of nuclei comes about in the following two ways. Firstly, a change in nuclear positions results in a change in the electrostatic interaction energy, specifically the electron-nuclear and the nuclear-nuclear electrostatic energy. Secondly, the wavefunctions and the electron-density, implicitly depend on the nuclear positions. However, the latter does not contribute to the ionic forces as the first variation of the electronic ground-state energy with respect to the wavefunctions and the electron density vanishes. Thus, in principle, in a continuous setting, the ionic force is simply the classical electrostatic force, which is the celebrated Hellman-Feynman theorem\cite{hellmann,Feynmanforces}. However, in practice, in a discrete setting, there is often a need to account for additional contributions known as Pulay forces~\cite{pulay1987analytical}, arising due to the dependence of the basis set on the positions of the nuclei. Similar considerations are needed while computing the stress tensor, which is the derivative of the electronic ground-state energy with respect to the strain tensor. As a consequence, there exists a large body of work describing the calculation  of nuclear forces\cite{ihm1979momentumcfforce,Bendt1983-om,Scheffler1985-xb,Nielsen1985-an, Fournier1989-fy,Soler1989-sl,Jackson1990-eb,Soler1990-sw,Yu1991-st, Delley1991-up,Goedecker1992-at,Savrasov1992-bu,Pople1992-uf,Methfessel1993-ej,Fahnle1995-jt,Kohler1996-oj,Kresse1996-ae,Wills2000-ig,Soler2002-mk,Alemany2004-vc,Miyazaki2004-sw,Blum2009-gr,Hine2011-gs,Ruiz-Serrano2012-aw,Zhang2017-gg} and stress tensor\cite{Slater1972-tv,Janak1974-tv,Nielsen1983-ve,Yin1983-dk,Nielsen1985-em,Nielsen1985-an,Dal_Corso1994-we,Kresse1999-yi,Kudin2000-vu,Soler2002-mk,Thonhauser2002-oi,Doll2004-db,Torrent2008-nw,Nagasako2011-yr,Knuth2015-fe,Sharma2018-gc,Becker2019-bt,belbase2021stress} in the discrete basis-sets used to solve the Kohn-Sham problem.

In this work, we derive the expressions for ionic forces and stress tensor in DFT calculations using the enriched finite element (EFE) basis, and conduct verification studies to ascertain the accuracy. The finite element (FE) basis, which comprises of local piecewise continuous polynomials as basis functions, offers systematic convergence and excellent parallel scalability in DFT calculations, given the local nature of the basis. While many prior efforts have developed and explored the use of FE basis for DFT calculations (cf.~e.g.~\cite{White1989, Tsuchida1996, Tsuchida1998, Pask1999, Pask2001, Pask2005, Zhang2008, Suryanarayana2010, Fang2012, Bao2012, Motamarri2013}), recent developments~\cite{Motamarri2020,Das2022} have demonstrated the utility of FE basis for conducting fast and accurate large-scale pseudopotential DFT calculations involving many tens of thousands of electrons~\cite{Das2019,Motamarri2020,NatNano2020,Das2022}. However, for all-electron electronic structure calculations, although prior works~\cite{White1989, Tsuchida1996, Batcho2000, Bylaska2009, Lehtovaara2009, Alizadegan2010, Bao2012, Motamarri2013, Schauer2013, Motamarri2014, Maday2014, Davydov2016, Kanungo2019, Ghosh2019, Ghosh2021} have demonstrated the systematic convergence of the FE basis, the computational cost remains high, given the large number of FE basis functions that are needed to accurately describe the all-electron wavefunctions. The EFE basis, which enriches the FE basis with compactly supported enrichment functions---such as, for e.g., those constructed from single-atom wavefunctions---can be used to significantly reduce the number of FE basis functions, and consequently improve the computational efficiency of all-electron calculations. 
While EFE basis has been employed for both all-electron calculations~\cite{Sukumar2009,enrichedBikash,Yamakawa2005,rufus21} as well as calculations involving hard pseudopotentials~\cite{Pask2017,pask2012linear}, in the present work we consider the framework of all-electron density functional theory.

The general approach to computing atomic forces or stresses in most numerical implementations relies on the outer variations of the Kohn-Sham energy functional with respect to the position of atoms (for computing forces) or the lattice vectors (for computing stress tensor). In the present work, however, we adopt a configurational force approach, which is based on inner variations of the Kohn-Sham energy functional. The configurational forces correspond to the generalized variational derivative of the Kohn-Sham energy functional with respect to the position of a material point $\bx$. The formulation is closely related to a recent work by Motamarri and Gavini~\cite{Motamarri2018}, who proposed the configurational force approach for the FE basis. In the present work, we extend the formulation of configurational forces to the enriched FE basis proposed in ~\cite{enrichedBikash,rufus21}, and derive the additional contributions that arise from the atom-centered enrichment functions. An advantage of the configurational force approach is that it provides a unified framework to compute ionic forces as well as cell stresses for geometry optimization.

We present numerical results that demonstrate the accuracy and efficacy of the proposed formulation to compute forces and cell stresses using the enriched FE basis for all-electron calculations.
We perform calculations on two molecular systems---carbon monoxide (CO) and sulfur trioxide (SO\textsubscript{3})---to demonstrate the applicability of the formulation to compute ionic forces in non-periodic systems. Further, we consider two periodic calculations, involving stress computations for the diamond 8-atom unit cell as well as ionic forces for silicon carbide (SiC) with a divacancy. In each case, we find convergence rates of close to $\mathcal{O}(h^{2p-1})$ with respect to the mesh size $h$ and the FE order $p$. Additionally, the ionic forces and stresses computed from finite-differencing the electronic ground-state energy were in excellent agreement with those obtained from the derived expression, thus ascertaining the accuracy and variational consistency of the expressions. Moreover, we also compare the obtained ionic forces and cell-stresses against other widely used codes, and demonstrate good agreement. 

The remainder of the paper is organized as follows. In Sec.~\ref{sec:form}, we derive expressions for the ionic forces and stress tensor for all-electron Kohn-Sham density functional theory calculations using the enriched FE basis. We do so by first presenting the real-space formulation of the Kohn-Sham variational problem employed in this work in Sec.~\ref{subsec:ksdft} and Sec.~\ref{subsec:rs_ksdft}. Subsequently, we provide details of the enriched finite element basis employed in this work in Sec.~\ref{subsec:efe}. The expressions of the configurational forces are derived in Sec.~\ref{subsec:gen_cf} for the enriched finite element basis. We conclude the section by describing the utility of configurational forces to evaluate ionic forces and the stress tensor in Sec.~\ref{sec:nucforstrtens}. Next, in Sec.~\ref{sec:res}, we present the numerical study and results that demonstrate the  accuracy and efficacy of the proposed formulation to compute ionic forces and cell stresses. Finally, in Sec.~\ref{sec:summary}, we summarize our findings and present an outlook.

\section{\label{sec:form}Formulation}
In this section, we derive the expressions for ionic forces and stress tensor in all-electron Kohn-Sham density functional theory calculations using the enriched FE basis. In particular, we use the configurational forces approach which facilitates the computation of both the ionic forces and cell stresses in a unified framework. A configurational force is simply the G\^{a}teaux derivative of the Kohn-Sham electronic ground-state energy functional along the direction of a prescribed deformation of the underlying space, with the deformation being characterized by a generator function. The configurational forces corresponding to an appropriate choice of generator functions are in turn used to evaluate the ionic forces and the stress tensor. The idea of using configurational forces to evaluate ionic forces and stress tensor in the context of FE basis was previously proposed for orbital-free DFT~\cite{das2015real} and Kohn-Sham DFT~\cite{Motamarri2018}. In this work, we extend the formulation to all-electron Kohn-Sham DFT calculations using enriched FE basis, which is a mixed basis constructed by augmenting the FE basis with atom-centered numerical enrichment functions. While the derivation presented in this work is similar in spirit to~\cite{Motamarri2018}, there are some key differences. Firstly, we consider the discrete Kohn-Sham energy functional as our starting point, as opposed to the continuous setting in which the configurational forces were derived in the previous work~\cite{Motamarri2018}. This is essential in order to account for, as well as delineate, the contributions from the enrichment functions to the configurational forces. In the absence of basis enrichment functions, the expressions derived in this work would reduce to the ones presented in~\cite{Motamarri2018}.
Secondly, in this work, we use the smeared charge approach proposed in~\cite{pask2012linear}, which enables a more efficient treatment of the electrostatic interactions arising from nuclear charges. 

In the following subsections, we present the formulation and derive the expressions for computing the ionic forces and stress tensor. For brevity, we only consider periodic systems. The expressions for the non-periodic case may be deduced by simply replacing the Brillouin zone sampling by an evaluation at $\bk=0$. In Sec.~\ref{subsec:ksdft} and \ref{subsec:rs_ksdft}, we present the Kohn-Sham energy functional and the real-space formulation used in this work. This is followed by a brief description of the enriched FE employed in this work in Sec.~\ref{subsec:efe}. We present the configurational forces in Sec.~\ref{subsec:gen_cf}, and present the approach to evaluate the ionic forces and stress tensor in Sec.~\ref{sec:nucforstrtens}.

\subsection{Kohn-Sham Density Functional Theory}\label{subsec:ksdft}
Consider a periodic unit cell $\Omega$ containing $N_e$ electrons and $N_a$ nuclei with ionic position vectors $\bR = \{\bR_1, \bR_2, \cdots ,\bR_{N_a} \}$. Neglecting spin, the free energy of the system in Kohn-Sham density functional theory~\cite{dftHohKohn,dftKohnSham} at finite temperature~\cite{merminFiniteTemp} is given by

\begin{equation}\label{eq_FreeEnergyFunctional}
\begin{split}
\mathcal{F}(\boldsymbol{\Gamma}^{\mathbf{u}_{\bk}}, \boldsymbol{\mathbf{u}_{\bk}}, \mathbf{R}) 
=T_{\mathrm{s}}(\boldsymbol{\Gamma}^{\mathbf{u}_{\bk}}, \boldsymbol{\mathbf{u}_{\bk}}, \mathbf{R})+&E_{\mathrm{xc}}(\rho)+E_{\mathrm{el}}(\rho, \mathbf{R})-\\
&E_{\mathrm{ent}}(\boldsymbol{\Gamma}^{\mathbf{u}_{\bk}}) \:,
\end{split}
\end{equation}
where $\mathbf{u}_{\bk} = \{ u_{1,\bk}(\bx), u_{2,\bk}(\bx),\cdots, u_{N,\bk}(\bx)\} \:\: (N>N_e /2)$ represent electronic wavefunctions corresponding to $\bk$, a point in the reciprocal space. These wavefunctions are considered to be non-orthogonal, in general. $\boldsymbol{\Gamma}^{\mathbf{u}_{\bk}}$ represents the matrix corresponding to the single-particle density operator ($\hat{\Gamma}$) expressed in the non-orthogonal basis $\mathbf{u}_{\bk}$. Hence, the elements of the matrix are given by $\Gamma_{pq}^{\mathbf{u}_{\bk}} = \sum_{r=1}^N {{S^{\bk}}^{-1}_{pr}} \langle u_{r,\bk}| \hat{\Gamma}| u_{q,\bk}\rangle $, where $S^{\bk}$ is the overlap matrix whose elements are given by
\begin{equation}\label{eq_WavefunctionOverlapS}
    S^{\bk}_{m,n} = \int_{\Omega} u_{m,\bk}^{*}(\bx) \,u_{n,\bk}(\bx) d\bx \:.
\end{equation}
The superscript `$^*$' in the above expression denotes complex conjugate. The electron density $\rho(\bx)$ in Eq.~(\ref{eq_FreeEnergyFunctional}) can be written in terms of the density matrix and non-orthogonal wavefunctions as follows:
\begin{equation}\label{eq_ElectronDensity}
    \rho(\mathbf{x})=2 \fint_{\Omega_{\text{BZ}}} \left\{ \sum_{p, q, r=1}^{N} \Gamma_{p q}^{\mathbf{u}_{\bk}} \,\, {{S^{\bk}}^{-1}_{q r}} u_{r,\bk}^{*}(\bx) \,u_{p,\bk}(\bx) \right\} d\bk \:, 
\end{equation}
where the average integral over the Brillouin zone, in practice, is computed as a discrete sum over $\bk$-points  lying in the Brillouin zone ($BZ$) with an associated weight  $w_{\bk}$. A common choice for the $k$-point grid is the Monkhorst-Pack (MP) scheme~\cite{monkhorst1976special}, which is adopted in this work. In the present work, we restrict our analysis to spin-independent systems. However, all the ideas discussed subsequently can be generalized to spin-dependent systems.

The first constituent of the free energy functional in Eq.~(\ref{eq_FreeEnergyFunctional}) is the kinetic energy of the non-interacting electrons, $T_s$, which is given by
\begin{widetext}
\begin{equation}\label{eq_KineticEnergy}
    T_{\text{s}}\left(\boldsymbol{\Gamma}^{\mathbf{u}_{\bk}}, \mathbf{u}_{\bk}\right)=2 \fint_{\Omega_{\text{BZ}}} \left\{\sum_{p, q, r=1}^{N} \int_{\Omega} \Gamma_{p q}^{\mathbf{u}_{\bk}} {S^{\bk}}^{-1}_{q r} u_{r,\bk}^{*}(\mathbf{x})\left(-\frac{1}{2} (\nabla + i\bk)^{2}\right) u_{p,\bk}(\mathbf{x}) d \mathbf{x} \right\} d\bk \:.
\end{equation}
\end{widetext}
The exchange-correlation functional in the local density approximation (LDA)\cite{ceperley1980ground,perdewzunger} adopted in this work is given by
\begin{equation} \label{eq_excorenergy}
    E_{\text{xc}}(\rho) = \int_{\Omega} F(\rho) d\bx = \int_{\Omega} \varepsilon_{xc} (\rho) \rho(\bx) d\bx \,.
\end{equation}
$E_{\text{el}}$ represents the classical electrostatic interaction energy between the electrons and nuclei and can be written as follows:
\begin{equation}\label{eq_ElectrostaticEnergy}
  E_{\text{el}}(\rho, \mathbf{R}) = E_{\text{H}}(\rho)  + E_{\text{ext}}(\rho, \mathbf{R})  + E_{\text{ZZ}}(\bR) \:,
\end{equation}
where $E_{\text{H}}$ and $E_{\text{ZZ}}$ are the Hartree energy and the nuclear repulsive energy, respectively, and are given by
\begin{equation}
    E_{\mathrm{H}}(\rho) =\frac{1}{2}  \int_{\Omega} \int_{\mathbb{R}^3} \frac{\rho(\bx) \rho(\bx^{\prime})}{|\bx-\bx^{\prime}|} d \bx^{\prime} d \bx \:,
\end{equation}
and 
\begin{equation}
    E_{\mathrm{zz}}(\bR)=\frac{1}{2} \sum_{I} \sum_{J,\,J\neq I}  \frac{Z_{I} Z_{J}}{\left|\mathbf{R}_{I}-\mathbf{R}_{J}\right|} \:.
\end{equation}
Note that, in the above equation, the sum over $J$ represents all nuclei in $\mathbb{R}^3$ while the sum over $I$ includes only those in the domain $\Omega$. $E_{\text{ext}}$ in Eq.~(\ref{eq_ElectrostaticEnergy}) represents the interaction between electrons and nuclei and is given by
\begin{equation}
    E_{\mathrm{ext}}(\rho,\bR)=-\sum_{J} \int_{\Omega} \rho(\mathbf{x}) \frac{Z_{J}}{\left|\mathbf{x}-\mathbf{R}_{J}\right|} d \mathbf{x} \:.
\end{equation}
Lastly, the entropic energy of the electrons in Eq.~(\ref{eq_FreeEnergyFunctional}) is given by
\begin{equation}
    E_{\text{ent}}(\boldsymbol{\Gamma}^{\mathbf{u}_{\bk}})=-2 \sigma\fint_{\Omega_{\text{BZ}}}\text{tr}\left[\boldsymbol{\Gamma}^{\mathbf{u}_{\bk}} \ln \boldsymbol{\Gamma}^{\mathbf{u}_{\bk}}+\left(\boldsymbol{I}-\boldsymbol{\Gamma}^{\mathbf{u}_{\bk}} \right) \ln \left(\boldsymbol{I}-\boldsymbol{\Gamma}^{\mathbf{u}_{\bk}}\right)\right] d\bk \:,
\end{equation}
where $\sigma = k_B T$ with $k_B$ denoting the Boltzmann constant and $T$ denoting the electronic temperature.

The electronic ground-state, for given nuclear positions, is given by the following variational problem:
\begin{equation} \label{eq_VaritionalProblemWithoutElectro}
     \min_{\boldsymbol{\Gamma}^{\mathbf{u}_{\bk}} \in \mathbb{R}^{N\times N}} 
     \min_{\boldsymbol{\mathbf{u}_{\bk}} \in (H_{\text{per}}^1(\Omega))^N}
     \mathcal{F}_{c}(\boldsymbol{\Gamma}^{\mathbf{u}_{\bk}},\boldsymbol{\mathbf{u}_{\bk}} , \mathbf{R}) \:, \quad \forall\, \bk \in BZ 
\end{equation}
where $\mathcal{F}_{c}$ is a constrained free-energy functional given by 
\begin{equation}
    \mathcal{F}_{c}=\mathcal{F}-\mu\left[2 \operatorname{tr}\left(\fint_{\Omega_{\text{BZ}}} \boldsymbol{\Gamma}^{\mathbf{u}_{\bk}}d \bk\right)-N_{e}\right] \:.
\end{equation}
In the above equation, $\mu$, which is the Lagrange multiplier corresponding to the constraint on the number of electrons, also denotes the Fermi level. We note that $H_{\text{per}}^1(\Omega)$ denotes the Hilbert space of periodic functions such that the functions and their first-order derivatives are square-integrable in $\Omega$.

\subsection{Local real-space formulation of Kohn-Sham DFT}\label{subsec:rs_ksdft}
Here we present the local real-space formulation that forms the basis for the derivation of the configurational forces expression. To begin with, the electrostatic interaction energy Eq.~(\ref{eq_ElectrostaticEnergy}) presented in the last section are extended in real-space. However, these extended interactions can be recast into a local variational formulation using auxiliary electrostatic potentials (cf.~\cite{das2015real}) as:
\begin{widetext}
\begin{equation}\label{eq_LocalElectrostaticsPointCharge}
\begin{split}
    E_{\mathrm{H}}(\rho) + E_{\mathrm{ext}}(\rho,\bR) + E_{\mathrm{zz}}(\bR) = \max_{\phi \in H_{\text{per}}^1(\Omega)} \left\{ \int_{\Omega}   \Big[ (\rho(\bx) + b(\bx,\bR))\phi(\bx) - 
    \frac{1}{8\pi} \left|\del \phi(\bx)\right|^2  \Big] d\bx \right\} - \\
     \sum_{I=1}^{N_a} \max_{V_{I} \in H^{1}\left(\mathbb{R}^{3}\right)} \left\{\int \Big[ b_I(\bx,\bR)  V_{I}(\bx) - \frac{1}{8 \pi} \left|\nabla V_{I}(\bx)\right|^{2} \Big] d \bx \right\}\:.
\end{split}
\end{equation}
\end{widetext}
In the above equation,
$b(\bx,\bR) = \sum_{I} b_I(\bx,\bR) = \sum_{I} -Z_{I} \tilde{\delta}\left(\left|\bx-\mathbf{R}_{I}\right|\right) $ is the nuclear charge density represented using regularized Dirac-delta functions centered at the nucleus. Also, the second term in Eq.~(\ref{eq_LocalElectrostaticsPointCharge}) serves the purpose of removing the nuclear self-interaction contribution contained in the first term. 

In the present work, however, we adopt the smeared charge representation for the nuclear charge~\cite{pask2012linear} which is computationally more efficient in a discrete setting. To this end, the nuclear charge density is written as
\begin{equation}\label{eq_SmearedCharge}
b_{s}(\bx) = \sum_{I} b_{s,I}(\bx) =   \sum_{I} -Z_{I} g(|\bx - \bR_{I}|, r_{c,I})\,,
\end{equation}
where $g(|\bx-\bR_I|,r_{c,I})$ denotes a smeared charge which is localized within $|\bx-\bR_I| < r_{c,I}$ and integrates to unity. We employ the following form for the unit smeared charge~\cite{pask2012linear}
\begin{equation} 
    g(r,r_c) = \begin{cases}
    \frac{-21 (r-r_c)^3 (6r^2 + 3r r_c + r_c^2)}{5 \pi r_c^8},  & 0\leq r \leq r_c, \\
    0, & r > r_c
    \end {cases}
\end{equation}
The $r_{c,I}$'s are chosen to be the largest possible values that avoid overlap between two neighboring smeared charges.
We note that, while not explicitly indicated, the smeared charge sphere is considered to wrap around the periodic boundary should it breach the boundaries of the computational domain. 
The local reformulation for the electrostatic energy using smeared nuclear charge is given by
\begin{equation}\label{eq_LocalElectrostaticsSmearedChargeFormulation}
\begin{split}
    E_{\text{el}}(\rho, \mathbf{R}) &= E_{\mathrm{H}}(\rho) + E_{\mathrm{ext}}(\rho,\bR) + E_{\mathrm{zz}}(\bR) \\
    &= \max_{\phi \in H_{\text{per}}^1(\Omega)} \mathcal{L}_{\text{el}}(\phi, \rho, \mathbf{R})\:,
\end{split}
\end{equation}
where
\begin{widetext}
\begin{equation}\label{eq_LocalElectrostaticsSmearedChargeFunctional}
\begin{split}
     \mathcal{L}_{\text{el}}(\phi, \rho, \mathbf{R})= &  \int_{\Omega}   \Big[ (\rho(\bx) + b_s(\bx,\bR))\phi(\bx) - 
    \frac{1}{8\pi} \left|\del \phi(\bx)\right|^2  \Big] d\bx  
    + \sum_I \intomegai \rho(\bx) \Big[ V_{I}\left(|\bx - \bR_I|\right) - 
     V_{s,I}\left(|\bx - \bR_I|,r_{c,I}\right) \Big] d\bx \\
     & - 
     \sum_I \frac{1}{2}\, \intomegai   b_{s,I}(\bx)    V_{s,I}\left(|\bx - \bR_I|,r_{c,I}\right) d\bx \:.
\end{split}
\end{equation}
\end{widetext}
In the above, $(\rho(\bx) + b_s(\bx,\bR))$ constitutes a charge-neutral system. The second term is a correction term containing the difference between the exact nuclear potential ($V_{I}$) corresponding to the point nuclear charge and the smeared nuclear potential ($V_{s,I}$) corresponding to $b_{s,I}$. These are given by
\begin{equation} 
    V_{I}(r) = -\frac{Z_I}{r}\,,
\end{equation}
\begin{equation} 
    V_{s,I}(r,r_{c,I}) = -Z_I v_g(r,r_{c,I})\,,
\end{equation}
where $v_g(r,r_c)$ is the potential corresponding to the $g(r,r_c)$ and is given by
\begin{equation} \label{eqnvg}
    v_g(r,r_c) = \begin{cases}
    \frac{9r^7 - 30r^6r_c + 28r^5r_c^2 - 14r^2r_c^5 +12r_c^7}{5r_c^8}, &0\leq r \leq r_c \\
    \frac{1}{r}, &r > r_c\,.
    \end{cases}
\end{equation}
We note that since $V_I$ and $V_{s,I}$ are identical for $r>r_{c,I}$, the correction for each nucleus is evaluated only inside the sphere of radius $r_{c,I}$ around $\bR_I$, whose domain is denoted by $\Omega_I$. The last term in Eq.~(\ref{eq_LocalElectrostaticsSmearedChargeFunctional}) removes the self-interaction energy corresponding to the smeared nuclear charges.

Finally, the electronic ground-state energy for a given positions of the nuclei $\bR$ can be computed using Eq.~(\ref{eq_VaritionalProblemWithoutElectro}) and Eq.~(\ref{eq_LocalElectrostaticsSmearedChargeFormulation}) as the following variational problem:
\begin{widetext}
\begin{equation}\label{eq_LocalGroundstateFormulation}
E_{0}(\mathbf{R}) = \mathcal{L}(\bar{\boldsymbol{\Gamma}}^{\mathbf{u}_{\bk}}, {\bar{\mathbf{u}}_{\bk}}, \bar{\phi}, \mathbf{R}) 
= \min_{\boldsymbol{\Gamma}^{\mathbf{u}_{\bk}} \in \mathbb{R}^{N\times N}} 
\min_{\boldsymbol{\mathbf{u}_{\bk}} \in (H_{\text{per}}^1(\Omega))^N} \max_{\phi \in H_{\text{per}}^1(\Omega)}
\mathcal{L}(\boldsymbol{\Gamma}^{\mathbf{u}_{\bk}}, \boldsymbol{\mathbf{u}_{\bk}},\phi, \mathbf{R}) \:,
\end{equation}
where
\begin{equation}\label{eq_LocalGroundstateFormulationFunctional}
    \mathcal{L}(\boldsymbol{\Gamma}^{\mathbf{u}_{\bk}}, \boldsymbol{\mathbf{u}_{\bk}},\phi, \mathbf{R}) = T_{\mathrm{s}}(\boldsymbol{\Gamma}^{\mathbf{u}_{\bk}}, \boldsymbol{\mathbf{u}_{\bk}}, \mathbf{R})+E_{\mathrm{xc}}(\rho)+\mathcal{L}_{\text{el}}(\phi, \rho, \mathbf{R}) -
    E_{\mathrm{ent}}(\boldsymbol{\Gamma}^{\mathbf{u}_{\bk}})-\mu\left[2 \operatorname{tr}\left(\fint_{\Omega_{\text{BZ}}}\boldsymbol{\Gamma}^{\mathbf{u}_{\bk}} d \bk \right) -N_{e}\right] \:.
\end{equation}
\end{widetext}
In the above, $\mathcal{L}$ denotes the local reformulation of the constrained free-energy functional $\mathcal{F}_c$, and $\bar{\boldsymbol{\Gamma}}^{\mathbf{u}_{\bk}}, \bar{\mathbf{u}}_{\bk}$ and $\bar{\phi}$ denote the extremizers of the variational problem.

\subsection{Enriched Finite Element basis}\label{subsec:efe}
The enriched finite element (EFE) basis used in this work is constructed by augmenting the finite element (FE) basis with atom-centered enrichment functions. The enrichment functions, which attempt to capture the highly oscillatory behavior of the wavefunctions around the nuclei, obviate the need for highly refined finite element mesh near the nuclei, thus greatly reducing the computational cost. In this work, we derive configurational forces in the context of enriched finite element basis presented by Kanungo and Gavini~\cite{enrichedBikash}. We note that using the EFE basis tends to yield ill-conditioned matrices in both the generalized eigenvalue problem and the linear system of equations (electrostatics). This affects the robustness of the self-consistent field iterations. There are many approaches~\cite{Schweitzer2011,Albrecht2018} proposed in recent years to address this problem. In particular, Rufus et. al.~\cite{rufus21} have proposed the orthogonalized enriched FE (OEFE) basis, in which, the atomic solutions are orthogonalized with respect to the underlying FE basis before augmenting as enrichments. The OEFE basis yields well-conditioned matrices ensuring numerical robustness. However, we note that the EFE basis presented in~\cite{enrichedBikash} and the OEFE basis presented in ~\cite{rufus21} span the same function space. Thus, in this work we employ the following strategy to compute ionic forces and cell stresses. First, we compute the electronic ground-state using the OEFE basis. We subsequently transform the fields from the OEFE to the EFE basis. Finally, we evaluate the forces and stresses using the expressions presented in this work. We derive the configurational force expressions in the EFE basis to avoid accounting for the additional contributions arising from the orthogonalizing terms in the OEFE basis.

Next, we provide a brief overview of the EFE basis. In the EFE basis, the wavefunctions $u_{\alpha,\bk}(\bx)$ are approximated as 
\begin{equation} \label{eqnUOEFE}
\begin{split}
\ualbk(\bx) \approx \,& \ualbk^h(\bx)\\
=&\underbrace{\sum_{i=1}^{n_h}\shp{C}{i}u_{\alpha,\bk,i}^C}_{\text{Classical FE}} + \underbrace{\sum_{I=1}^{N_a}\sum_{j=1}^{n_I} \shpR{E,u_{\bk}}{j,I}{\bx-\bR_I} u_{\alpha,\bk,j,I}^{E}}_{\text{Enrichment}}\,.
\end{split}
\end{equation}
In the above equation, the superscript $h$ indicates a discrete field, and the superscript $C$ and $E$ are used to distinguish the classical and the enrichment components, respectively. In particular, $\shp{C}{i}$ denotes the $i\textsuperscript{th}$ classical FE basis function, and $u_{\alpha,\bk,i}^C$ denotes the expansion coefficient of $\shp{C}{i}$ for $\ualbk$. Similarly, $\shp{E,u_{\bk}}{j,I}$ denotes the $k$-point dependent enrichment function for $\ualbk~(\forall \alpha)$. The index $I$ runs over all the atoms ($N_a$) in the system, and the index $j$ runs over all the atomic Kohn-Sham orbitals ($n_I$) we include for the atom $I$. In other words, the $I$\textsuperscript{th} atom, situated at $\bR_I$, contributes $n_I$ enrichment functions, each centered around $\bR_I$. $u_{\alpha,\bk,j,I}^E$ represents the expansion coefficient of $\shpR{E,u_{\bk}}{j,I}{\bx-\bR_I}$  corresponding to $\ualbk$.  The reader is referred to~\cite{enrichedBikash,rufus21} for the form of $\shpR{E,u_{\bk}}{j,I}{\bx-\bR_I}$. 

In a similar vein, the electrostatic potential  $\phi$ is given as  
\begin{equation} \label{eqnPhiEFE}
    \phi(\bx) \approx \phi^h(\bx)=\underbrace{\sum_{j=1}^{n_h}\shp{C}{j}\phi_{j}^C}_{\text{Classical FE}} + \underbrace{\sum_{I=1}^{N_a}\shpR{E,\phi}{I}{\bx - \bR_I}\phi_{I}^E}_{\text{Enrichment}}\,.
\end{equation}
As with the discretization of $\ualbk$ (Eq.(~\ref{eqnUOEFE})), $\shp{C}{j}$ denotes the $j\textsuperscript{th}$ classical FE basis function and $\phi_j^C$ denotes its corresponding coefficient. Similarly, $\shpR{E,\phi}{I}{\bx - \bR_I}$ is the $I\textsuperscript{th}$  enrichment function 
with a corresponding coefficient $\phi_I^C$.  

Thus, in the finite-dimensional EFE basis, the electronic ground-state energy is given by 

\begin{widetext}
\begin{equation}
\begin{split}
E_0^h(\bR) 
   &=\mathcal{L}\bigg(
        \bar{\boldsymbol{\Gamma}}^{\mathbf{u}_{\bk}}, 
        \bigg\{ 
            u^h_{\alpha,\bk}\Big(
                \{\shp{C}{i}\},
                \{\bar{u}_{\alpha,\bk,i}^C\},
                \{\shpR{E,u_{\bk}}{j,I}{\bx-\bR_I}\},
                \{\bar{u}_{\alpha,\bk,j,I}^{E}\}
            \Big) 
        \bigg\}, 
        \phi^h \Big(
            \{\shp{C}{j}\},
            \{\bar{\phi}_{j}^C\},
            \{\shpR{E,\phi}{I}{\bx - \bR_I}\},
            \{\bar{\phi}_{I}^E\}
        \Big), \mathbf{R}
    \bigg) \\
   &= \min_{\boldsymbol{\Gamma}^{\mathbf{u}_{\bk}} \in      \mathbb{R}^{N\times N}} 
    \min_{ \{u_{\alpha,\bk,i}^C\}\in \mathbb{C}^{N\times{n_h}}}
    \min_{ \{u_{\alpha,\bk,\nu}^E\}\in \mathbb{C}^{N\times{n_{E}^{u_{\bk}}}}} 
    \max_{\{\phi_{j}^C\} \in \mathbb{R}^{n_h} }
    \max_{\{\phi_{j}^E\} \in \mathbb{R}^{N_a} }  \\
   &\,\,\mathcal{L}\bigg(
        \boldsymbol{\Gamma}^{\mathbf{u}_{\bk}}, 
        \bigg\{ 
            u^h_{\alpha,\bk}\Big(
                \{\shp{C}{i}\},
                \{u_{\alpha,\bk,i}^C\},
                \{\shpR{E,u_{\bk}}{j,I}{\bx-\bR_I}\},
                \{u_{\alpha,\bk,j,I}^{E}\}
            \Big) 
        \bigg\},
        \phi^h \Big(
            \{\shp{C}{j}\},
            \{\phi_{j}^C\},
            \{\shpR{E,\phi}{I}{\bx - \bR_I}\},
            \{\phi_{I}^E\}
        \Big), \mathbf{R} \:
    \bigg)   \:.
\end{split}
\end{equation}
\end{widetext}
In the above equation, $\{\,.\,\}$ are used to denote lists of functions or scalar coefficients running over all indices. $\bar{\boldsymbol{\Gamma}}^{\mathbf{u}_{\bk}},\{\bar{u}_{\alpha,\bk,i}^C\}, \{\bar{u}_{\alpha,\bk,j,I}^E\} ,\{\bar{\phi}_{j}^C\} \text{ and } \{\bar{\phi}_{I}^E\}$ denote the discrete ground-state solution (stationarity point of $\mathcal{L}$) for given nuclear positions $\bR$ and basis functions. These quantities can be computed by solving the Kohn-Sham equations in the EFE basis. In particular, in the context of canonical eigenfunctions, $\{\bar{u}_{\alpha,\bk,i}^C\}$  and  $\{\bar{u}_{\alpha,\bk,\nu}^E\}$ are the solution of the generalized Kohn-Sham eigenvalue problem; $\bar{\boldsymbol{\Gamma}}^{\mathbf{u}_{\bk}}$ contains the fractional occupancies of the eigenstates given by the Fermi-Dirac distribution; $\{\bar{\phi}_{j}^C\}  \text{ and } \{\bar{\phi}_{I}^E\}$ are solutions to the Poisson problem governing the electrostatic interactions.

\subsection{Configurational forces}\label{subsec:gen_cf}

We now derive expressions for the configurational force corresponding the electronic ground-state free energy in Eq.~(\ref{eq_LocalGroundstateFormulation}). We employ the approach of inner variations, where the generalized force corresponding to a perturbation of underlying space is evaluated. To this end, we define a bijective mapping $\boldsymbol{\tau}^\varepsilon \,:  \mathbb{R}^3 \mapsto \mathbb{R}^3 $ which represents an infinitesimal perturbation of the underlying space, mapping a material point $\bx$ to a new point $\bx^{\mathbf{\varepsilon}} = \boldsymbol{\tau}^\varepsilon(\bx) $. Using a Taylor series expansion in $\varepsilon$, this mapping can be written as
\begin{equation}\label{eqn_deformationtau}
    \boldsymbol{\tau}^\varepsilon(\bx) = \bx + \varepsilon \boldsymbol{\Upsilon}(\bx) + \mathcal{O}(\varepsilon^2) \: ,
\end{equation}
where $\boldsymbol{\Upsilon}(\bx) = \left. \frac{d}{d\varepsilon} \boldsymbol{\tau}^\varepsilon(\bx)\right \vert_{\varepsilon=0}$ is the generator of the spatial perturbation. We note that $\bx^{0} = \bx$ denotes the reference configuration.

The configurational force corresponding to a prescribed  $\boldsymbol{\tau}^\varepsilon(\bx)$ is given by
\begin{equation}\label{eqn_fconstdiscrete}
\left. \frac{d}{d\varepsilon} 
\mathcal{L}^\varepsilon(\densOpMatEpsBar,\wfnGrpAtKEpsBar,\phiEpsBar,\mathbf{R}^\varepsilon)
\right \vert_{\varepsilon=0}\,,
\end{equation}
where $\densOpMatEpsBar$, $\wfnGrpAtKEpsBar$ and $\phiEpsBar$ are the extremizers of constrained free energy functional in the perturbed configuration ($\mathcal{L}^\varepsilon$). In other words, the configurational force is the G\^ateaux derivative of $\mathcal{L}^\varepsilon$ in the direction of $\boldsymbol{\Upsilon}(\bx)$ with all the electronic fields set to the electronic ground-state.

In the discrete setting with the enriched finite element basis, the configurational force is given by

\begin{widetext}
\begin{equation}\label{eqn:cfdef}
\begin{split}
\hat{F}^h(\boldsymbol{\Upsilon}(\bx)) =  \frac{d}{d\varepsilon} 
\mathcal{L}^{\varepsilon}
    \bigg(
        \densOpMatEpsBar,
        \bigg\{
            \wfnTildeDiscNumKEvalAt
                {\alpha}
                {\bk}
                {\collectionOf{\shpFuncClassEps{i}}}
                {\collectionOf{\coeffUClassicalEpsBar{\alpha}{i}}}
                {\collectionOf{\shpFuncUEnrichmentEps{j}{I}}}
                {\collectionOf{\coeffUEnrichmentEpsBar{\alpha}{j}{I}}}
        \bigg\},\\
        \phiTildeDiscEvalAt
                {\collectionOf{\shpFuncClassEps{j}}}
                {\collectionOf{\coeffPhiClassicalEpsBar{j}}}
                {\collectionOf{\shpFuncPhiEnrichmentEps{I}}}
                {\collectionOf{\coeffPhiEnrichmentEpsBar{I}}}, 
        \mathbf{R^\varepsilon}
    \bigg)
 \bigg|_{\varepsilon=0}\,.      
\end{split}
\end{equation}
\end{widetext}
We note that in computing the configurational force in the discrete setting, we choose the generator from the subspace spanned by the FE basis (cf.~Sec~\ref{sec:nucforstrtens}) as it allows us to take advantage of the isoparametric nature of FE basis functions. In the above, $\densOpMatEpsBar, \collectionOf{\coeffUClassicalEpsBar{\alpha}{i}}, \collectionOf{\coeffUEnrichmentEpsBar{\alpha}{j}{I}} , \collectionOf{\coeffPhiClassicalEpsBar{j}} $ and $\collectionOf{\coeffPhiEnrichmentEpsBar{I}}$ represent the ground-state electronic fields in the discrete setting, i.e., the extremizers of $\mathcal{L}^{\varepsilon}$ in subspace spanned by the enriched FE basis. It is important to note here that the enrichment functions $\{ N_{j,I}^{E,u_{\bk}}(\bx^\varepsilon - \bR^\varepsilon)\} $ and $\{ N^{E,\phi}_{I}(\bx^{\varepsilon} - \bR^\varepsilon)\}$ retain their form even when the underlying space is deformed, as these are \emph{a priori} computed enrichment functions independent of the finite element discretization.

For the sake of further simplification, we express the constrained free energy functional in terms of the energy density $f$ as
\begin{widetext}
\begin{equation}
\begin{split}
&\mathcal{L}\label{eq:fdef}
    \bigg(
        \densOpMat,
        \bigg\{
            \wfnTildeDiscNumKEvalAt
                {\alpha}
                {\bk}
                {\collectionOf{\shpFuncClass{i}}}
                {\collectionOf{\coeffUClassical{\alpha}{i}}}
                {\collectionOf{\shpFuncUEnrichment{j}{I}}}
                {\collectionOf{\coeffUEnrichment{\alpha}{j}{I}}}
        \bigg\},
        \phiTildeDiscEvalAt
                {\collectionOf{\shpFuncClass{j}}}
                {\collectionOf{\coeffPhiClassical{j}}}
                {\collectionOf{\shpFuncPhiEnrichment{I}}}
                {\collectionOf{\coeffPhiEnrichment{I}}}, 
        \mathbf{R}
    \bigg)   \\
&= \int_{\Omega} 
   f\bigg(
        \densOpMat,
        \bigg\{
            \wfnTildeDiscNumKEvalAt
                {\alpha}
                {\bk}
                {\collectionOf{\shpFuncClass{i}}}
                {\collectionOf{\coeffUClassical{\alpha}{i}}}
                {\collectionOf{\shpFuncUEnrichment{j}{I}}}
                {\collectionOf{\coeffUEnrichment{\alpha}{j}{I}}}
        \bigg\},
        \phiTildeDiscEvalAt
                {\collectionOf{\shpFuncClass{j}}}
                {\collectionOf{\coeffPhiClassical{j}}}
                {\collectionOf{\shpFuncPhiEnrichment{I}}}
                {\collectionOf{\coeffPhiEnrichment{I}}}, 
        \mathbf{R}
    \bigg) d\bx \:.
\end{split}    
\end{equation}

The configurational force is decomposed as follows (refer to Appendix for details):

\begin{equation}\label{eq_cfsplit}
\begin{split}
\hat{F}^h(\boldsymbol{\Upsilon}(\bx)) 
&=
\frac{d}{d\varepsilon} \Bigg[ \int_{\Omega^\varepsilon} f^{\varepsilon} \bigg(
 \densOpMatEpsBar,
        \bigg\{
            \wfnTildeDiscNumKEvalAt
                {\alpha}
                {\bk}
                {\collectionOf{\shpFuncClassEps{i}}}
                {\collectionOf{\coeffUClassicalEpsBar{\alpha}{i}}}
                {\collectionOf{\shpFuncUEnrichmentEps{j}{I}}}
                {\collectionOf{\coeffUEnrichmentEpsBar{\alpha}{j}{I}}}
        \bigg\},\\
        &\qquad \qquad \qquad
        \phiTildeDiscEvalAt
                {\collectionOf{\shpFuncClassEps{j}}}
                {\collectionOf{\coeffPhiClassicalEpsBar{j}}}
                {\collectionOf{\shpFuncPhiEnrichmentEps{I}}}
                {\collectionOf{\coeffPhiEnrichmentEpsBar{I}}}, 
        \mathbf{R^\varepsilon}    
    \bigg) d\bx^\varepsilon  \Bigg] \bigg|_{\varepsilon=0} \\
&=
\hat{F}_{1}(\boldsymbol{\Upsilon}(\bx)) + \hat{F}_{2}(\boldsymbol{\Upsilon}(\bx)) \:,
\end{split}  
\end{equation}
where $\hat{F}_{1}$ denotes the contribution to the configurational force that arise from all dependencies on $\varepsilon$ excepting those that arise from the enrichment functions, and $\hat{F}_{2}$ denotes the contribution to the configurational force that arise from the enrichment functions. In particular,    

\begin{equation}
\label{eq:hatF_1}
\begin{split}
 \hat{F}_{1}(\boldsymbol{\Upsilon}(\bx)) 
 &=
 \frac{d}{d\varepsilon} \Bigg[ \int_{\Omega^\varepsilon} f^{\varepsilon}\bigg(
 \densOpMatEpsBar,
        \bigg\{
            \wfnTildeDiscNumKEvalAt
                {\alpha}
                {\bk}
                {\collectionOf{\shpFuncClass{i}}}
                {\collectionOf{\coeffUClassicalEpsBar{\alpha}{i}}}
                {\collectionOf{\shpFuncUEnrichment{j}{I}}}
                {\collectionOf{\coeffUEnrichmentEpsBar{\alpha}{j}{I}}}
        \bigg\},\\
        &\qquad \qquad \qquad
        \phiTildeDiscEvalAt
                {\collectionOf{\shpFuncClass{j}}}
                {\collectionOf{\coeffPhiClassicalEpsBar{j}}}
                {\collectionOf{\shpFuncPhiEnrichment{I}}}
                {\collectionOf{\coeffPhiEnrichmentEpsBar{I}}}, 
        \mathbf{R^\varepsilon}    
    \bigg) d\bx^\varepsilon  \Bigg] \bigg|_{\varepsilon=0} \,.
\end{split}    
\end{equation}
\end{widetext}
We note that, in writing the above expression, we have made use of the isoparametric nature of the FE basis functions, i.e.,     $\shpFuncClassEps{i} = \shpFuncClass{i}$. We note that the configurational force is of a similar form as the one considered in Motamarri \& Gavini~\cite{Motamarri2018}, albeit in a discrete setting. Thus, using the previously derived results~\cite{Motamarri2018}, we note that we can evaluate $\hat{F}_{1}(\boldsymbol{\Upsilon}(\bx))$ in terms of the canonical Kohn-Sham eigenfunctions as follows:
\begin{widetext}
\begin{equation}
\begin{split}
 \hat{F}_{1}(\boldsymbol{\Upsilon}(\bx))  =  \int_{\Omega} \mathbf{E}: \nabla \boldsymbol{\Upsilon}(\mathbf{x}) d \mathbf{x} + \mathrm{F}^{K} + \mathrm{F}^{\text{S}}\:,
\end{split}
\end{equation}
where the Eshelby tensor $\mathbf{E}$ is given by 
\begin{equation}\label{eqn_cff1}
\begin{split}
\mathbf{E} 
= 
\Bigg(
    \fint_{\Omega_{\text{BZ}}} 
    \Bigg[  
        \sum_{\alpha} \bar{f}_{\alpha,\bk}   \: 
            \bigg\{ 
                \del \, \wfnDiscNumKConjBar{\alpha}{\bk}(\bx)  \cdot \del \wfnDiscNumKBar{\alpha}{\bk}(\bx) \,  -
                2i \, \wfnDiscNumKConjBar{\alpha}{\bk}(\bx) \bk\cdot\del \, \wfnDiscNumKBar{\alpha}{\bk}(\bx)  +
                (|\bk|^2 -2  \epsalbkBar) |\wfnDiscNumKBar{\alpha}{\bk}(\bx)|^2 
            \bigg\}   
    \Bigg] 
    d\bk \:  \\
   + \varepsilon_{xc} (\rhoDiscBar) \rhoDiscBar(\bx) + 
    (\rhoDiscBar(\bx) + b_s(\bx,\bR))\hbarr{\phi}(\bx) - 
    \frac{1}{8\pi} \big| \del \hbarr{\phi}(\bx ) \big|^2 
\Bigg) \boldsymbol{I}  \\
- 
\fint_{\Omega_{\text{BZ}}} 
\Bigg[ 
    \sum_{\alpha} \bar{f}_{\alpha,\bk}   \: 
    \bigg\{  
        \del \, \wfnDiscNumKConjBar{\alpha}{\bk}(\bx) \otimes \del \,  \wfnDiscNumKBar{\alpha}{\bk}(\bx) + 
        \del \, \wfnDiscNumKBar{\alpha}{\bk}(\bx) \otimes \del \, \wfnDiscNumKConjBar{\alpha}{\bk}(\bx) - 
        2 i \, \wfnDiscNumKConjBar{\alpha}{\bk}(\bx) \big(\del \,  \wfnDiscNumKBar{\alpha}{\bk}(\bx) \otimes \bk \big) 
    \bigg\} 
\Bigg] 
d\bk  \\
 +\frac{1}{4\pi} \del \hbarr{\phi}(\bx) \otimes \del \hbarr{\phi}(\bx) \:,
\end{split}
\end{equation}

and 
\begin{equation}
\mathrm{F}^{K} 
 = 
\fint_{\Omega_{\text{BZ}}} 
\Bigg[ 
    \sum_{\alpha} \bar{f}_{\alpha,\bk}  \int_{\Omega} \: 
    \bigg\{  
        -2i \, \wfnDiscNumKConjBar{\alpha}{\bk}(\bx) \left. \frac{d}{d\varepsilon} \bk^\varepsilon \right \vert_{\varepsilon=0} \cdot\del \,  \wfnDiscNumKBar{\alpha}{\bk}(\bx) + 
        \left. \frac{d}{d\varepsilon} |\bk^\varepsilon|^2   \right \vert_{\varepsilon=0}|\wfnDiscNumKBar{\alpha}{\bk}(\bx)|^2\bigg\} d\bx 
\Bigg] 
d\bk \:.
\end{equation}
We note that, in the above expression, $\bk^\varepsilon$ depends on the nature of the generator. For a Gaussian-type generator used in the force calculations (cf. Sec.~\ref{sec:nucforstrtens}), since there is no change in the lattice vectors, we have
\begin{equation}
    \bk^\varepsilon(\bk) = \bk \implies 
    \left. \frac{d}{d\varepsilon} \bk^\varepsilon \right \vert_{\varepsilon=0} = \mathbf{0}
\end{equation}
However, under affine deformations, this dependence is given by:
\begin{equation}
    \bk^\varepsilon(\bk) = (\mathbf{I} - \varepsilon \mathbf{C}^T)\bk
    \implies 
    \left. \frac{d}{d\varepsilon} \bk^\varepsilon \right \vert_{\varepsilon=0} =  -\mathbf{C}^T\bk \:,
\end{equation}
where $\mathbf{C}$ is a second order tensor independent of $\bx$.\newline
$\mathrm{F}^{\text{S}}$ in Eq.(\ref{eqn_cff1}) contains contributions from the smeared nuclear charges, and is given by
\begin{equation}
\begin{split}
 \mathrm{F}^{\text{S}} = \sum_I  \Bigg[\intomegai  \hbarr{\rho}(\bx)   \del \Big( V_{I}\left(|\bx - \bR_I|\right)  - V_{s,I}\left(|\bx - \bR_I|,r_{c,I}\right) \Big) \cdot \Big( \boldsymbol{\Upsilon}(\bx) - \boldsymbol{\Upsilon}(\bR_I)  \Big) d\bx \Bigg]    \\
 +\sum_I \Bigg[\intomegai   \hbarr{\rho}(\bx) \Big\{  V_{I}\left(|\bx - \bR_I|\right) - V_{s,I}\left(|\bx - \bR_I|,r_{c,I}\right) \Big\} \del \cdot \boldsymbol{\Upsilon}(\bx) d\bx \Bigg]  \\ 
 - \sum_I \Bigg[\intomegai  \hbarr{\phi}(\bx) Z_{I} \del g(|\bx - \bR_{I}|, r_{c,I}) \cdot \Big( \boldsymbol{\Upsilon}(\bx) - \boldsymbol{\Upsilon}(\bR_I)  \Big)  d\bx \Bigg] \:.
\end{split}
\end{equation}

We now turn our attention to $\hat{F}_{2}(\boldsymbol{\Upsilon}(\bx))$ in Eq.(\ref{eq_cfsplit}), which contains the contributions arising from the $\varepsilon$-dependence of the enrichment functions. We provide here the expression for $\hat{F}_{2}(\boldsymbol{\Upsilon}(\bx))$, and refer to the Appendix for the details of the derivation:

\begin{equation}\label{eq_finalF2}
\begin{split}
 \hat{F}_{2}(\boldsymbol{\Upsilon}(\bx)) =  & \fint_{\Omega_{\text{BZ}}} \Bigg[  \sum_{\alpha} \bar{f}_{\alpha,\bk}  \int_{\Omega} \: \bigg\{ \del \, \Big( \left.  \frac{d}{d\varepsilon}  \widetilde{\ualbk^{h,\varepsilon \: *}}(\bx^\varepsilon)  \right \vert_{\varepsilon=0} \Big) \cdot \del \, \wfnDiscNumKBar{\alpha}{\bk}(\bx) + 
       \del \,  \wfnDiscNumKConjBar{\alpha}{\bk}(\bx)  \cdot \del \, \Big( \left. \frac{d}{d\varepsilon}\widetilde{\ualbk^{h,\varepsilon}}(\bx^\varepsilon) \right \vert_{\varepsilon=0} \Big)  \\
     & - 2i \, \epsder{\widetilde{\ualbk^{h,\varepsilon \: *}}(\bx^\varepsilon)} \bk\cdot\del \,  \wfnDiscNumKBar{\alpha}{\bk}(\bx) - 
       2i \, \wfnDiscNumKConjBar{\alpha}{\bk}(\bx) \bk\cdot\del \, \epsder{\widetilde{\ualbk^{h,\varepsilon}}(\bx^\varepsilon)} + |\bk|^2  \epsder{\widetilde{\ualbk^{h,\varepsilon \: *}}(\bx^\varepsilon)} \wfnDiscNumKBar{\alpha}{\bk}(\bx)  \\
     & \qquad \qquad +|\bk|^2   \wfnDiscNumKConjBar{\alpha}{\bk}(\bx) \epsder{\widetilde{\ualbk^{h,\varepsilon }}(\bx^\varepsilon)} \bigg\}   d\bx\Bigg] d\bk  \\
     & +\intomega  \left. \ddeps \widetilde{\rho^{h,\varepsilon}} (\eps{\bx}) \right \vert_{\varepsilon=0} \big[ V_{\text{xc}}(\bx) + \hbarr{\phi}(\bx) + \sum_I  ( V_{I}(|\bx-\bR_I|) - V_{s,I}(|\bx-\bR_I|) )\big] d\bx  \\
     & \qquad \qquad +\intomega   (\hbarr{\rho}(\bx) + b_s(\bx,\bR)) \ddeps \widetilde{\phi^{h,\varepsilon}} (\eps{\bx}) \evalepszero d\bx  - \intomega \frac{1}{4\pi}     \del \big( \ddeps \widetilde{\phi^{h,\varepsilon}}(\eps{\bx})\big) \evalepszero \cdot \del \hbarr{\phi}(\bx)    d\bx  \\
      & - \fint_{\Omega_{\text{BZ}}} \Bigg[  \sum_{\alpha} \bar{f}_{\alpha,\bk}  \int_{\Omega} \:  2  \epsalbkBar \bigg\{ \epsder{\widetilde{\ualbk^{h,\varepsilon \:*}}(\bx^\varepsilon)} \wfnDiscNumKBar{\alpha}{\bk}(\bx) + 
     \wfnDiscNumKConjBar{\alpha}{\bk}(\bx) \epsder{\widetilde{\ualbk^{h,\varepsilon}}(\bx^\varepsilon)} \bigg\}   d\bx\Bigg] d\bk \:,
\end{split}
\end{equation}
where 
\begin{equation} \label{eq_utilde_epsder}
\begin{split}
    \wfnSpecialEpsDer
    &\coloneqq 
    \left. \frac{d}{d\varepsilon} \wfnTildeDiscNumKEvalAt
                {\alpha}
                {\bk}
                {\collectionOf{\shpFuncClass{i}}}
                {\collectionOf{\coeffUClassicalBar{\alpha}{i}}}
                {\collectionOf{\shpFuncUEnrichmentEps{j}{I}}}
                {\collectionOf{\coeffUEnrichmentBar{\alpha}{j}{I}}}
    \right \vert_{\varepsilon=0} \\
    &=
    \sum_{I=1}^{N_a}\sum_{j=1}^{n_I} \bar{u}_{\alpha,\bk,j,I}^{E} \: \del \shpR{E,u_{\bk}}{j,I}{\bx-\bR_I} \cdot \Big( \boldsymbol{\Upsilon}(\bx) - \boldsymbol{\Upsilon}(\bR_I)  \Big) \: 
\end{split}
\end{equation}
and 
\begin{equation} \label{eq_phitilde_epsder}
\begin{split}
    \phiSpecialEpsDer 
    &\coloneqq 
    \ddeps
        \phiTildeDiscEvalAt
                    {\collectionOf{\shpFuncClass{j}}}
                    {\collectionOf{\coeffPhiClassicalBar{j}}}
                    {\collectionOf{\shpFuncPhiEnrichmentEps{I}}}
                    {\collectionOf{\coeffPhiEnrichmentBar{I}}}   
    \evalepszero \\
    &=\sum_{I=1}^{N_a} \bar{\phi}_{I}^E \: \del \shpR{E,\phi}{I}{\bx - \bR_I} \cdot \Big( \boldsymbol{\Upsilon}(\bx) - \boldsymbol{\Upsilon}(\bR_I)  \Big)
\end{split}
\end{equation}
\end{widetext}
account for contributions from the enrichment functions. We note that $\epsder{\widetilde{\rho^{h,\varepsilon}}(\bx^\varepsilon)}$ can be computed using Eq.~(\ref{eq_utilde_epsder}). We remark that 
$\hat{F}_{2}(\boldsymbol{\Upsilon}(\bx))$ vanishes when the enrichment functions are absent, i.e., when $\bar{u}_{\alpha,\bk,j,I}^{E} = \bar{\phi}_{I}^E = 0$.

\subsection{\label{sec:nucforstrtens} Computation of ionic forces and stress tensor}

We now discuss the approach to compute the ionic forces and stress tensor from the configurational force. In order to compute the force  on a nucleus along the i\textsuperscript{th} direction, we choose a generator whose i\textsuperscript{th} component is a function centered at the nucleus with all other components to be zero. For instance, to compute the force on the $\text{I}^{\text{th}}$ nucleus in the x-direction, we consider the following generator 
\begin{equation}\label{eqn_forcgen}
    \boldsymbol{\Upsilon}(\bx) =  \begin{Bmatrix} f(\bx) \\ 0 \\ 0 \end{Bmatrix} \:.
\end{equation}
In particular, as the FE basis functions can also be used to perturb the space, owing to the isoparametric nature of the basis functions, we choose generators constructed from the FE basis. To this end, we use the lowest order subspace of the typically higher-order finite element function space employed. In other words, we choose $f$ to be given by
\begin{equation}
    f(\bx) = \sum_{i=1}^{8}\shp{C,\text{lin}}{i}  c_i \:,
\end{equation}
where $\shp{C,\text{lin}}{i}$ are trilinear FE basis functions which constitute a subspace of the original classical FE basis spanned by functions $\shp{C}{i}$.
The coefficients, $c_i$'s, are computed to be the value of $f$ at the corner vertices of finite element cells. The typical form of $f$ we employ in this work is a Gaussian centered at the nucleus whose ionic force we are interested in computing. Thus, 
\begin{equation}
    c_k = f(\bx_k) = g(|\bx_k - \bR_I|)\,. 
\end{equation}
We note that $g(r)$ is chosen such that it satisfies the following constraints: a) the support of the function is such that it does not contain any other nuclei other than the one under consideration, and b) $g(0) = 1$. The former constraint is essential to ensure that the sole contribution to the configurational force is from the nucleus of interest. The latter constraint mimics the perturbation of the underlying space displacing the  $\text{I}^{\text{th}}$ nucleus by $\varepsilon$ (Eq.~(\ref{eqn_deformationtau})). In the present work we choose
\begin{equation}
    g(r) = \exp(\alpha r^\beta) \:,
\end{equation}
with $\alpha = -0.8$ and $\beta = 4$. The force on the $\text{I}^{\text{th}}$ nucleus in the x-direction is simply the negative of the configurational force evaluated using the above generator (Eq.~(\ref{eqn_forcgen})). More importantly, in the context of structural relaxation, the value of the computed force and the nature of the generator determine not only the next position of the given nucleus, but also the FE mesh around it. 
 
In order to compute the stress tensor associated with cell relaxations, the prescribed perturbations correspond to an affine deformation of the periodic domain $\Omega_{p}$, which preserves the periodicity of the cell. To elaborate, the generator is given by $\boldsymbol{\Upsilon}(\bx) = \mathbf{C}\bx$, where $\mathbf{C}$, a second order tensor, is independent of $\bx$. By writing the stress tensor in terms of the derivative of the energy density with respect to strain tensor, and expanding the energy density in terms of the strain, it can be shown that~\cite{Motamarri2018}
    
\begin{equation} \label{eqn_stress}
\left. \frac{d}{d\varepsilon} \mathcal{F}_c^\varepsilon(\bar{\mathbf{u}}^\varepsilon,\bar{\mathbf{f}}^\varepsilon,\bar{\phi}^\varepsilon,\bR^\varepsilon) \right \vert_{\varepsilon=0} 
=
\Omega_p \frac{1}{2} \left( \left( \mathbf{C} + \mathbf{C}^{\text{T}}\right) : \boldsymbol{\sigma} \right)\,,
\end{equation} 
where $\boldsymbol{\sigma}$ is the stress tensor whose individual components can be computed by appropriately selecting the components of $\mathbf{C}$. For instance, to compute $\sigma_{\text{xx}}$, we choose $\mathbf{C}$ to be
\begin{equation*}
    \text{C} = \begin{Bmatrix} 1 & 0 & 0 \\ 0 & 0 & 0 \\  0 & 0 & 0 \end{Bmatrix}
    \;.
\end{equation*}

\section{\label{sec:res}Results and Discussion}

In this section, we present  numerical results that demonstrate the accuracy of the proposed formulation to compute forces and stresses using the enriched FE basis for all-electron calculations.
We first consider two molecular systems, carbon monoxide (CO) and sulfur trioxide ($\text{SO}_3$), to demonstrate the applicability of the formulation to non-periodic systems. This is followed by two periodic calculations, diamond 8-atom cubic cell and silicon carbide (SiC) cubic cell with a divacancy. For each of these benchmark systems, we perform the following studies. Firstly, we study the rate of convergence of the ionic force or stress tensor component with respect to discretization, i.e., decreasing finite element mesh size. We also benchmark the accuracy of our calculations with Gaussian basis for non-periodic systems and the linearized augmented planewave with local orbitals (LAPW+lo) basis for periodic systems. 
In the next study, we use the finite difference test wherein the underlying space is deformed by an infinitesimal amount using a generator and the force or the stress component is computed by finite-difference of the electronic ground-state energy at each configuration. This is compared against the values obtained using the configurational force expression. This finite difference test serves as a verification of the derived expressions, as well as establishes the variationality of the expression.
Finally, to further ascertain the variationality of the computed ionic force or stress tensor component, we vary the position of the nucleus or the lattice vector and study the agreement between the ionic force/stress tensor obtained by fitting the electronic ground-state energy to a higher order polynomial and the corresponding value evaluated using the configurational force expression. 

In all the following simulations, we use an n-stage Anderson mixing~\cite{Anderson1965} for density mixing with a stopping criterion of $10^{-8}$ on the $L_2$ norm of the density difference. Moreover, we set the electronic temperature, $T$, to 500K.

\subsection{\label{subsec:nonper} Non-periodic systems}

\subsubsection{\label{subsubsec:CO} Carbon monoxide}
\begin{figure}[htp]
    \centering
    \includegraphics[scale=0.1]{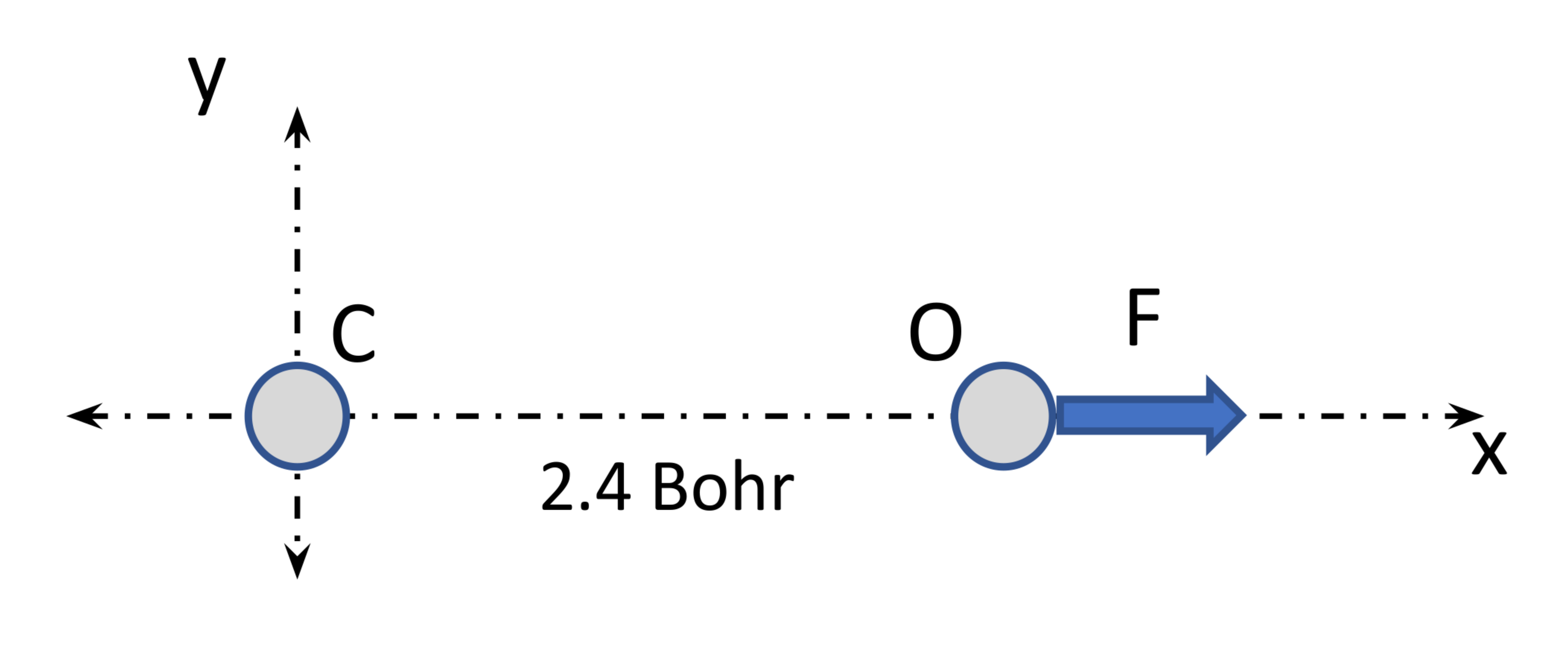}
    \caption{Schematic: CO molecule of bond length 2.4 Bohr~.}
    \label{fig:CO_convergence_schematic}
\end{figure}

\begin{figure}[htp]
    \centering
    \includegraphics[scale=1.1]{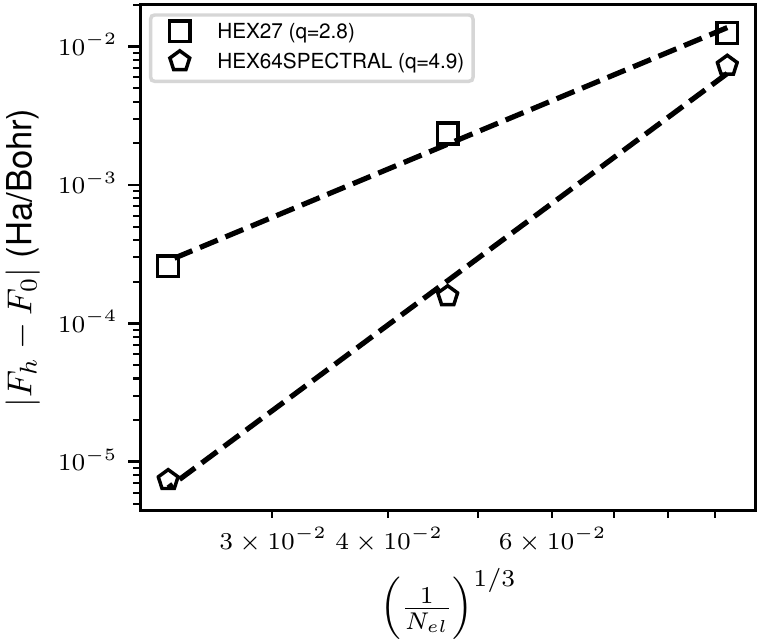}
    \caption{Convergence of force on the O atom of CO molecule.}
    \label{fig:CO_convergence}
\end{figure}

We consider a CO molecule of bond length 2.4 Bohr (cf. Fig.~\ref{fig:CO_convergence_schematic}) in a simulation box of 80 Bohr. We consider the x-component of the force acting on the O atom as metric to ascertain the accuracy. 
In order to demonstrate the convergence with respect to discretization, for a given set of enrichment functions, we compute the force $F_h$ using three levels of increasingly refined finite element meshes. We conduct this study using quadratic and cubic spectral finite elements. Figure~\ref{fig:CO_convergence} plots the error in the force against $(1/N_{el})^{1/3}$, where $N_{el}$ represents the number of finite elements. We note that $(1/N_{el})^{1/3}$ provides a measure of the finite element mesh size for convergence studies. The exact value of force, $F_0$, is computed by performing the same calculation using an EFE basis containing a highly refined quartic FE mesh. We find the $F_0$ to be 0.202802 Ha/Bohr. The corresponding force value using a pc-4 Gaussian basis in NWChem is 0.202830 Ha/Bohr. The error in the force for finite element discretization is expected to be of the form,
\begin{equation}
    |F_h - F_0| = C \left(\frac{1}{N_{el}}\right)^{q/3}\,,
\end{equation}
with $q$ denoting the rate of convergence. We note that, for both the element types, we find the value of $q$ to be close to $2p-1$, which is the optimal rate of convergence. 
Next, we conduct the finite difference test by applying the perturbations (Eq.~(\ref{eqn_deformationtau})) to the underlying space based on a generator (Eq.~(\ref{eqn_forcgen})) centered at the O atom by setting the $\varepsilon$ to $-0.02$, $-0.01$, $0.0$, $0.01$, and $0.02$. In each case, we perform the ground-state calculation to obtain the electronic ground-state energy. Using the energy values, we obtain the ionic force on the O atom in the x-direction, by using a 5-point stencil. We find that the finite-difference value and the configurational force agree to within $2.5 \times 10^{-6}$ Ha/Bohr.

\begin{figure}[htp]
    \centering
    \includegraphics[scale=1.0]{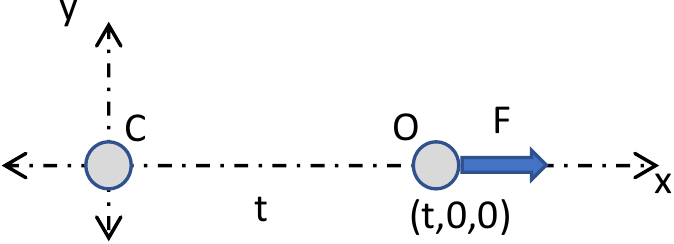}
    \caption{Schematic: CO variationality test}
    \label{fig:CO_variationality_schematic}
\end{figure}

In order to demonstrate the variationality of the computed ionic force, we keep the C atom fixed and vary the x-coordinate of the O atom as shown in Fig.~\ref{fig:CO_variationality_schematic}. In each case, we compute the electronic ground-state energy and the ionic force, $F$, on the O atom in the x-direction. 
\begin{figure}[htp]
    \centering
    \includegraphics[scale=1.1]{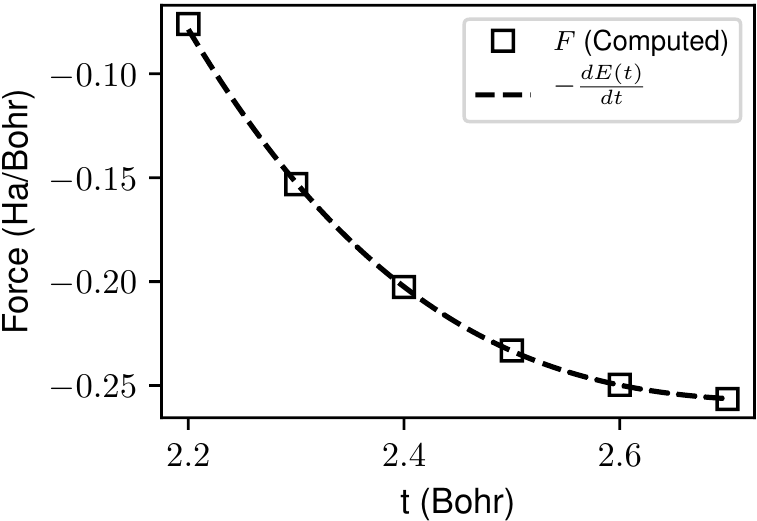}
    \caption{Comparison of computed force $F$  on the O atom and the derivative of energy fit $E(t)$ for CO.}
    \label{fig:CO_variationality}
\end{figure}
Fitting the energy using a fourth-order polynomial, we compare the force obtained using the derivative of the energy with that obtained from the configurational force expression in Fig.~\ref{fig:CO_variationality}. This agreement between the calculated ionic force and the force deduced from the energy ascertains the variational consistency of the formulation.

\subsubsection{\label{subsubsec:SO3} Sulfur trioxide}
\begin{figure}[htp]
    \centering
    \includegraphics[scale=0.08]{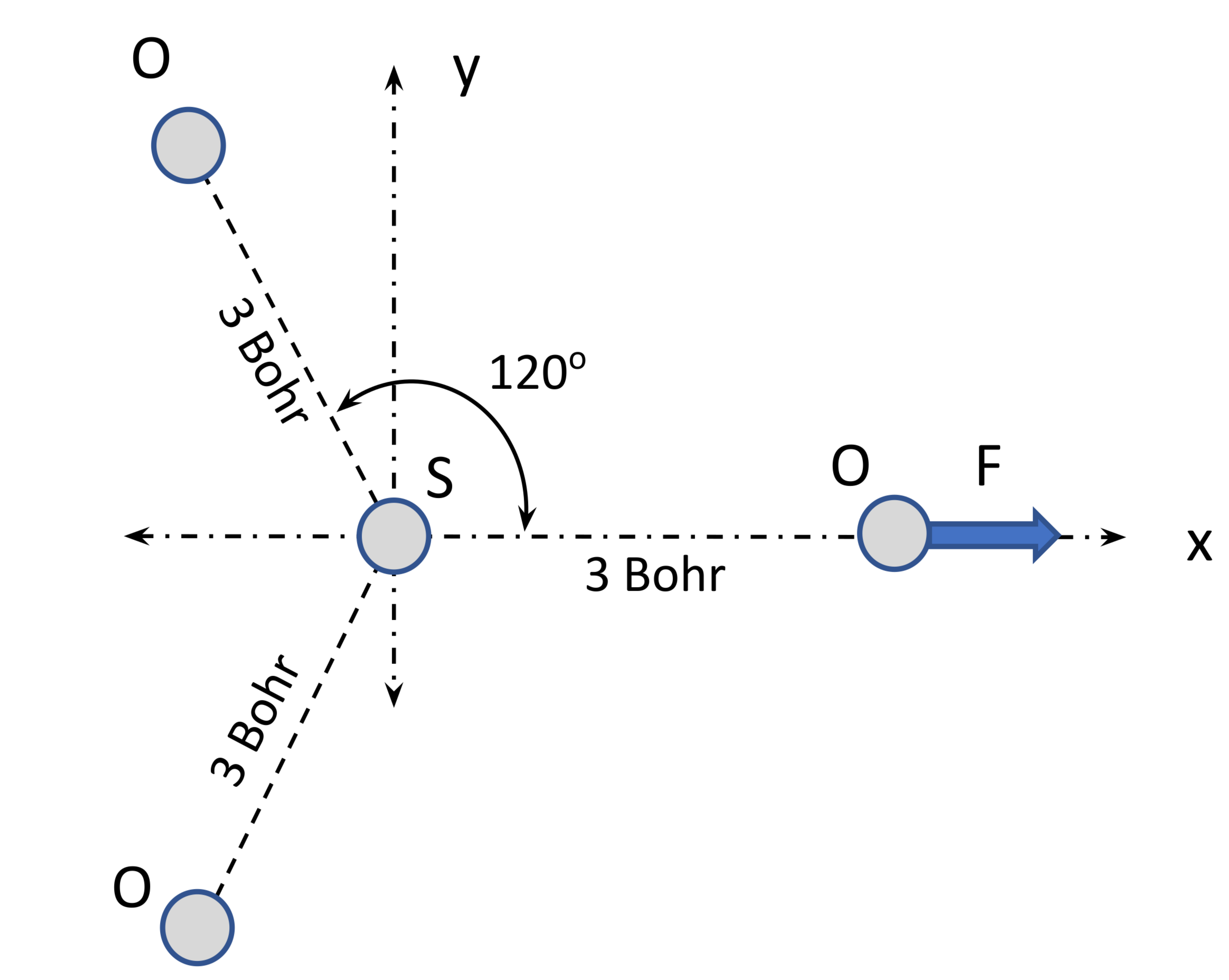}
    \caption{Schematic: SO\textsubscript{3} molecule of bond length 3 Bohr~.}
    \label{fig:SO3_convergence_schematic}
\end{figure}

We consider the planar SO\textsubscript{3} molecule of bond length 3 Bohr (cf. Fig.~\ref{fig:SO3_convergence_schematic}) in a simulation box of 80 Bohr. We consider the x-component of the force acting on the O atom as the metric of interest. 
\begin{figure}[htp]
    \centering
    \includegraphics[scale=1.1]{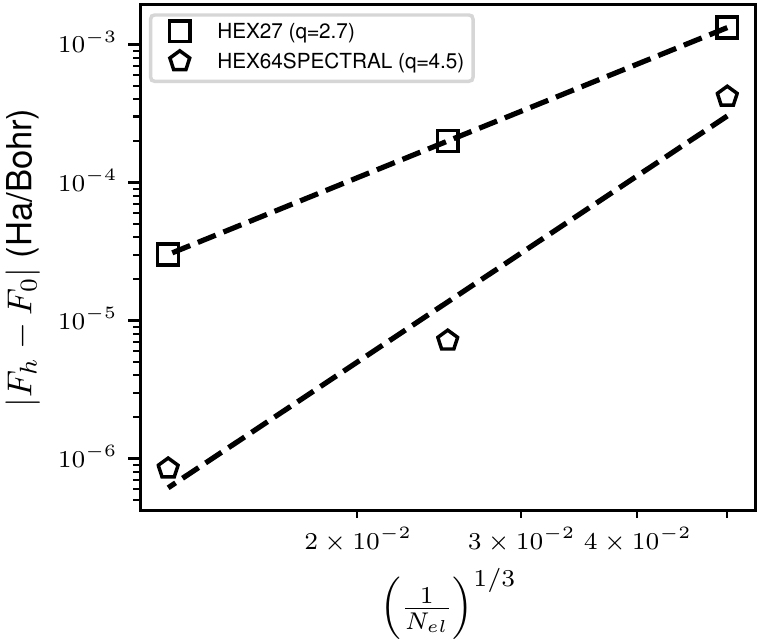}
    \caption{Convergence of force on the O atom for SO\textsubscript{3}~.}
    \label{fig:SO3_convergence}
\end{figure}
For a given set of enrichment functions, we compute the force, $F_h$, using three levels of increasingly refined finite element meshes, using quadratic and cubic spectral finite elements. 
Figure~\ref{fig:SO3_convergence} plots the error in the force against $(1/N_{el})^{1/3}$.
The value of $F_0$ is computed by performing the same calculation using an enriched basis containing a highly refined quartic FE mesh, which is computed to be 0.121280 Ha/Bohr. The corresponding force value using a pc-3 Gaussian basis (performed using NWChem) is 0.121051 Ha/Bohr. We note that the more accurate pc-4 basis calculations could not be performed due to SCF non-convergence arising from linear dependence in the basis.
 The rates of convergence, $q=2.7$ and $q=4.5$ for quadratic and cubic finite elements, respectively, are close to the optimal rates of convergence. 

\begin{figure}[htp]
    \centering
    \includegraphics[scale=1.1]{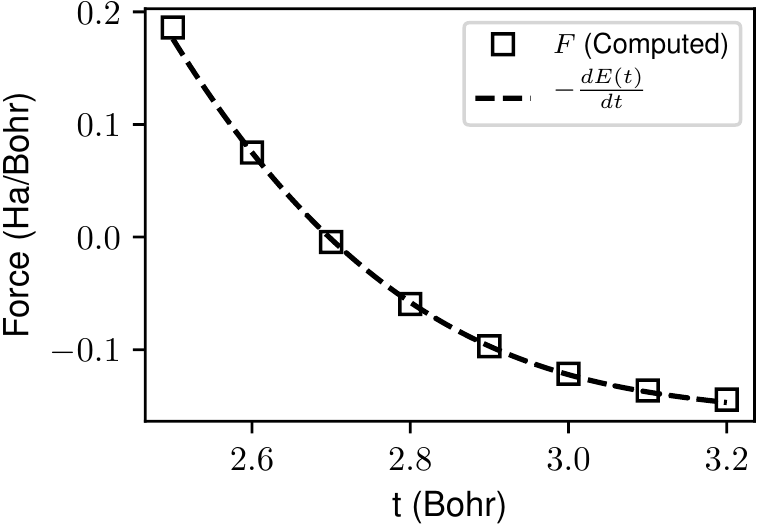}
    \caption{Comparison of computed force $F$ on the O atom and derivative of the energy fit $E(t)$ for SO\textsubscript{3}~.}
    \label{fig:SO3_variationality}
\end{figure}

The finite-difference test by perturbing the underlying space, in a manner similar to the CO molecule, is in agreement with the configurational force to within $3.5 \times 10^{-6}$ Ha/Bohr. Further, Fig.~\ref{fig:SO3_variationality} demonstrates the variationality by varying the x-coordinate of the O atom lying on the x-axis.
Again, we find an excellent agreement between the computed force and the force deduced from the energy. 

\subsection{\label{subsec:per} Periodic systems}
\subsubsection{\label{subsubsec:SiC-divac-force} Silicon carbide periodic cell with a divacancy}
We consider a silicon carbide cubic diamond-structured periodic cell with vacancies introduced as shown in Table ~\ref{table:SiCDivacCoord}. The quantity of interest we consider here is the force on the first C atom in the x-direction. All calculations performed for this material system involve sampling the Brillouin zone at the $\Gamma$-point. 
As with the previous two systems, we perform the convergence study using quadratic and cubic spectral finite elements. 
 
\begin{figure}[htp]
    \centering
    \includegraphics[scale=1.1]{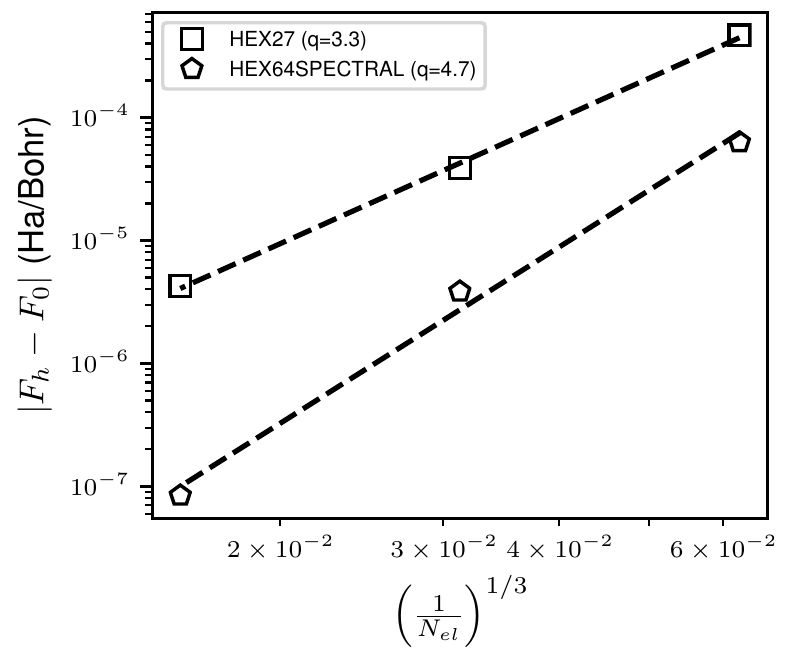}
    \caption{Convergence of the force on the C atom of SiC divacancy.}
    \label{fig:SiC_convergence}
\end{figure}

Figure~\ref{fig:SiC_convergence} plots the error in the force against $(1/N_{el})^{1/3}$. As evident from the figure, close to optimal rates of convergence are obtained for both the element types. The exact value of force $F_0$ is computed by performing the same calculation using an enriched basis containing a highly refined quartic FE mesh. We find the $F_0$ to be  0.0237091 Ha/Bohr. The corresponding force value obtained using the LAPW+lo basis in the Elk code is 0.0237962 Ha/Bohr. We note that this value was obtained using a low muffin-tin radius of 1.0 Bohr, as opposed to 1.8-2.0 Bohr typically used for the energy calculations. 
The spherical averaging inside the muffin-tins may have a more pronounced effect on the forces as compared to the energies. Calculations with muffin-tin radius lower than 1.0 Bohr were not performed in this study due to the steep increase in the cost associated with reducing the muffin-tin radius.

\begin{table}
\begin{tabular}{c | c  }
Species  & Fractional Coordinates  \\
\hline \hline
C & (0.0,0.0,0.0)  \\ 
C & (0.5,0.5,0.0)  \\
C & (0.0,0.5,0.5)  \\
x & (0.5,0.0,0.5)  \\ 
Si & (0.25,0.25,0.25)  \\
Si & (0.75,0.75,0.25)  \\
Si & (0.75,0.25,0.75)  \\
x & (0.25,0.75,0.75)  
\end{tabular}
\caption{SiC Divacancy fractional coordinates. x denotes vacancy site}
\label{table:SiCDivacCoord}
\end{table}

\begin{figure}[h]
    \centering
    \includegraphics[scale=1.1]{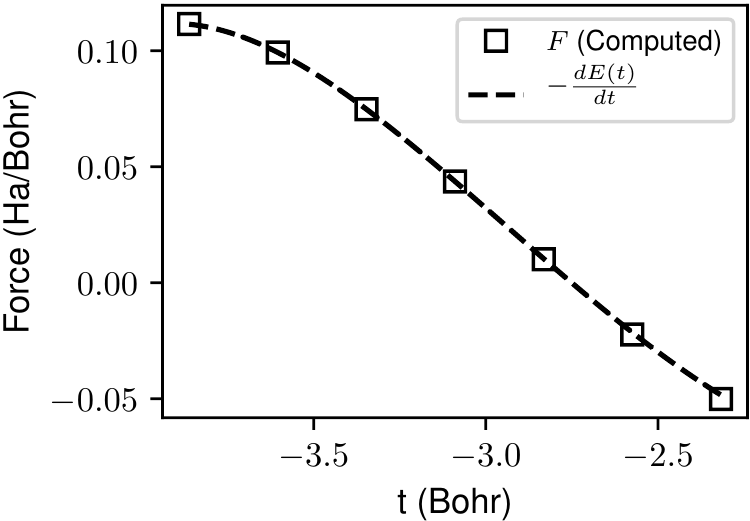}
    \caption{Comparison of computed force $F$ and  derivative of the energy fit $E(t)$ for SiC.}
    \label{fig:SiC_variationality}
\end{figure}

In the the finite difference test, we observe that the computed force and the finite-difference force agree to within $3.3 \times 10^{-6}$ Ha/Bohr. Next, we vary the z-coordinate of the first C atom to study the variationality of the computed force. The comparison between the calculated force and the force obtained using the derivative of the energy is shown in Fig.~\ref{fig:SiC_variationality}. In the figure, t is used to denote the z-coordinate of the first C atom.

\subsubsection{\label{subsubsec:C8-stress} Diamond unit cell}
We consider an 8-atom diamond unit cell of lattice constant $a =7.0$ Bohr. The quantity of interest we consider for this system is the hydrostatic stress. We first study the convergence rates of the stress with respect to the mesh size. We perform this study using quadratic and cubic spectral finite elements. For these calculations, the Brillouin zone is sampled at $\bk=(0.2,0.3,0.4)$ to verify the accuracy of the derived expressions and implementation when the computed KS wavefunctions are complex-valued.

\begin{figure}[htpb]
    \centering
    \includegraphics[scale=1.1]{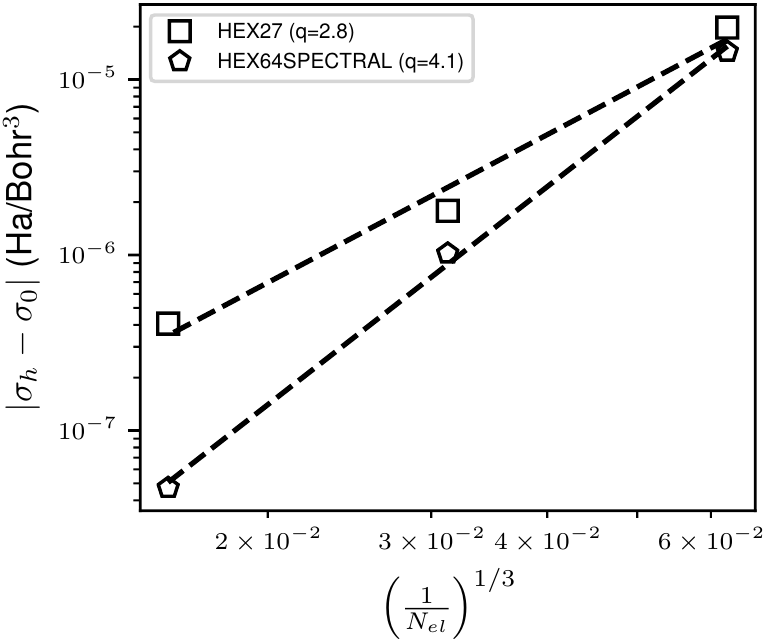}
    \caption{Convergence of hydrostatic stress for diamond unit cell.}
    \label{fig:C8_convergence}
\end{figure}

Figure~\ref{fig:C8_convergence} plots the error in the force against $(1/N_{el})^{1/3}$. 
As evident from the figure, close to optimal rates of convergence are obtained for both the element types. An enriched FE  calculation using a highly refined quartic mesh yielded $\sigma_0 = 0.00155146$ Ha/Bohr\textsuperscript{3}. The corresponding value of stress obtained using the LAPW+lo basis is $0.00154724$ Ha/Bohr\textsuperscript{3}. We note that this value was obtained by performing a finite-difference on the energy since the Elk code, used for benchmarking, does not currently have the capability to directly evaluate the value of stress. 

\begin{figure}[hbt!]
    \centering
    \includegraphics[scale=1.1]{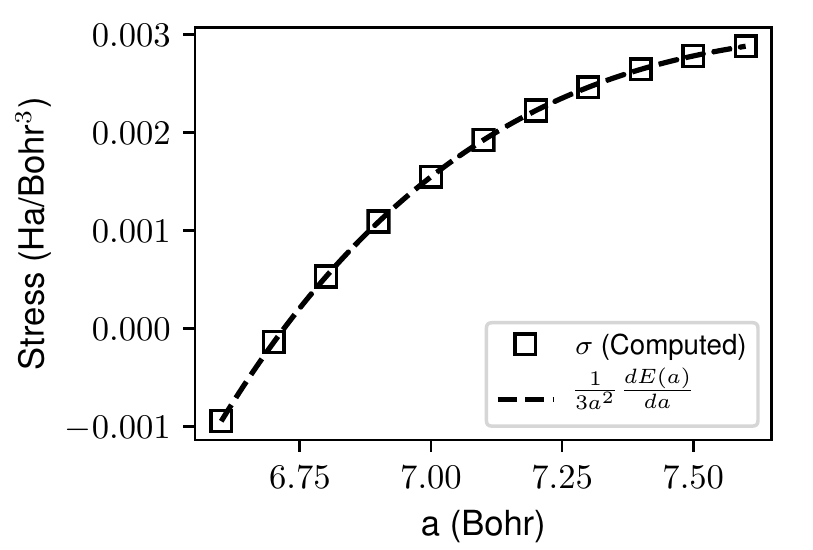}
    \caption{Comparison of computed stress $\sigma$ \& derivative of energy fit $E(t)$ for diamond.}
    \label{fig:C8_variationality}
\end{figure}

In order to conduct the finite difference test, we first deduce the value of hydrostatic stress from the electronic ground-state energies at lattice constants $a , a (1 \pm \varepsilon ), a (1 \pm 2\varepsilon ) $ using $\varepsilon = 0.01$. This value is then compared with the hydrostatic stress evaluated using configurational forces, and we find an agreement of $1.1 \times 10^{-6}$ Ha/Bohr\textsuperscript{3}. Finally, to study the variationality of the computed hydrostatic stress, we compare the computed value of stress using configurational forces at various lattice constants with the stress obtained by fitting the the electronic ground-state energy to a polynomial of the lattice constant. The comparison is shown in Fig.~\ref{fig:C8_variationality}, and we observe good agreement ascertaining the variational nature of the formulation.

\section{Summary} \label{sec:summary}
In the present work, we derived and implemented the configurational force approach in the context of enriched finite element basis to compute ionic forces and the stress tensor in all-electron density functional theory calculations. The approach provides a unified expression for both ionic forces and stress tensor to conduct structural relaxations. The derived configurational force is variational, and inherently accounts for Pulay corrections arising from the dependence of the basis on the nuclear positions. Further, both periodic and non-periodic calculations can be handled using the same framework. 

The accuracy of the formulation was verified using the four benchmark systems. Force calculations were demonstrated for CO and SO\textsubscript{3} molecules, and SiC periodic cell with a divacancy. Stress calculations were demonstrated for the diamond unit cell. In each case, we found convergence rates of $\mathcal{O}(h^{\alpha})$, $\alpha\approx 2p-1$, with respect to mesh size $h$, where $p$ denotes the polynomial order of the finite element basis. The finite difference test for each system showed a tight agreement between the force deduced from the energy and that evaluated using the derived configurational force expression. Moreover, for all systems we demonstrated the variationality of the proposed approach by showing a good agreement between the evaluated configurational force and the derivative of the polynomial fit of the electronic ground-state energy. 

Previous works~\cite{enrichedBikash,rufus21} have shown the merits of enriched finite elements for all-electron DFT calculations, such as systematic convergence, numerical efficiency and scalability. This work extends the utility of the enriched finite element basis by presenting an approach to compute the ionic forces and stress tensor for performing structural relaxations in all-electron DFT calculations. 

\acknowledgements
We gratefully acknowledge the support from the Department of Energy, Office of Basic Energy Sciences, grant number DE-SC0017380, under the auspices of which this work was conducted. V.G. also acknowledges the support from Toyota Research Institute that supported later parts of this work. This research used resources of the National Energy Research Scientific Computing Center, a DOE Office of Science User Facility supported by the Office of Science of the U.S. Department of Energy under Contract No. DE-AC02-05CH11231. V.G. also acknowledges the support of the Army Research Office through the DURIP grant W911NF1810242, which also provided the computational resources for this work.


\begin{widetext}

\section{Appendix}\label{sec:appendix}
We present the details of the derivation of the configurational force expression given in Sec.~\ref{sec:form}. The notation used here is consistent with the main text.
We recall from Eq.(\ref{eqn:cfdef}) that the configurational force is defined as 
\begin{equation}
\begin{split}
\hat{F}^h(\boldsymbol{\Upsilon}(\bx)) =  \frac{d}{d\varepsilon} 
\mathcal{L}^{\varepsilon}
    \bigg(
        \densOpMatEpsBar,
        \bigg\{
            \wfnTildeDiscNumKEvalAt
                {\alpha}
                {\bk}
                {\collectionOf{\shpFuncClassEps{i}}}
                {\collectionOf{\coeffUClassicalEpsBar{\alpha}{i}}}
                {\collectionOf{\shpFuncUEnrichmentEps{j}{I}}}
                {\collectionOf{\coeffUEnrichmentEpsBar{\alpha}{j}{I}}}
        \bigg\},\\
        \phiTildeDiscEvalAt
                {\collectionOf{\shpFuncClassEps{j}}}
                {\collectionOf{\coeffPhiClassicalEpsBar{j}}}
                {\collectionOf{\shpFuncPhiEnrichmentEps{I}}}
                {\collectionOf{\coeffPhiEnrichmentEpsBar{I}}}, 
        \mathbf{R^\varepsilon}
    \bigg)
 \bigg|_{\varepsilon=0}\,.      
\end{split}
\end{equation}
Since $\densOpMatEpsBar$, $\collectionOf{\coeffUClassicalEpsBar{\alpha}{i}}$, $\collectionOf{\coeffUEnrichmentEpsBar{\alpha}{j}{I}}$,$\collectionOf{\coeffPhiClassicalEpsBar{j}}$ and $\collectionOf{\coeffPhiEnrichmentEpsBar{I}}$ constitute the stationarity point of the constrained free-energy functional $\mathcal{L}^{\varepsilon}$, the partial derivative of $\mathcal{L}^{\varepsilon}$ with respect to any of these quantities vanishes. Hence, the configurational force expression reduces to 

\begin{equation}
\begin{split}
\hat{F}^h(\boldsymbol{\Upsilon}(\bx)) =  \frac{d}{d\varepsilon} 
\mathcal{L}^{\varepsilon}
    \bigg(
        \densOpMatBar,
        \bigg\{
            \wfnTildeDiscNumKEvalAt
                {\alpha}
                {\bk}
                {\collectionOf{\shpFuncClassEps{i}}}
                {\collectionOf{\coeffUClassicalBar{\alpha}{i}}}
                {\collectionOf{\shpFuncUEnrichmentEps{j}{I}}}
                {\collectionOf{\coeffUEnrichmentBar{\alpha}{j}{I}}}
        \bigg\},\\
        \phiTildeDiscEvalAt
                {\collectionOf{\shpFuncClassEps{j}}}
                {\collectionOf{\coeffPhiClassicalBar{j}}}
                {\collectionOf{\shpFuncPhiEnrichmentEps{I}}}
                {\collectionOf{\coeffPhiEnrichmentBar{I}}}, 
        \mathbf{R^\varepsilon}
    \bigg)
 \bigg|_{\varepsilon=0}\,,      
\end{split}
\end{equation}
where $\densOpMatBar$, $\collectionOf{\coeffUClassicalBar{\alpha}{i}}$, $\collectionOf{\coeffUEnrichmentBar{\alpha}{j}{I}}$,$\collectionOf{\coeffPhiClassicalBar{j}}$ and $\collectionOf{\coeffPhiEnrichmentBar{I}}$ are computed in the unperturbed ($\varepsilon=0$) configuration.

Next, we turn our attention to the basis functions. We note that, under a spatial perturbation, the form of the classical finite element basis function changes owing to the deformation of the underlying finite element mesh. However, the form of the enrichment functions, which are \emph{a priori} constructed, remains the unchanged. As noted in the main text, given the isoparametric nature of the FE basis functions, we have
\begin{equation}
    \shpFuncClassEps{i} = \shpFuncClass{i}\,.
\end{equation}
Using the notion of the energy density $f$ defined in Eq.(\ref{eq:fdef}), the free-energy functional in the deformed configuration is given by

\begin{equation}\label{eq:fdefInDeformed}
\begin{split}
\mathcal{L}^{\varepsilon}
    \bigg(
        \densOpMatBar,
        \bigg\{
            \wfnTildeDiscNumKEvalAt
                {\alpha}
                {\bk}
                {\collectionOf{\shpFuncClass{i}}}
                {\collectionOf{\coeffUClassicalBar{\alpha}{i}}}
                {\collectionOf{\shpFuncUEnrichmentEps{j}{I}}}
                {\collectionOf{\coeffUEnrichmentBar{\alpha}{j}{I}}}
        \bigg\},\\
        \phiTildeDiscEvalAt
                {\collectionOf{\shpFuncClass{j}}}
                {\collectionOf{\coeffPhiClassicalBar{j}}}
                {\collectionOf{\shpFuncPhiEnrichmentEps{I}}}
                {\collectionOf{\coeffPhiEnrichmentBar{I}}}, 
        \mathbf{R^\varepsilon}
  \bigg)\\
=  \int_{\Omega^\varepsilon} f^\varepsilon
        \bigg(
     \densOpMatBar,
        \bigg\{
            \wfnTildeDiscNumKEvalAt
                {\alpha}
                {\bk}
                {\collectionOf{\shpFuncClass{i}}}
                {\collectionOf{\coeffUClassicalBar{\alpha}{i}}}
                {\collectionOf{\shpFuncUEnrichmentEps{j}{I}}}
                {\collectionOf{\coeffUEnrichmentBar{\alpha}{j}{I}}}
        \bigg\},\\
        \phiTildeDiscEvalAt
                {\collectionOf{\shpFuncClass{j}}}
                {\collectionOf{\coeffPhiClassicalBar{j}}}
                {\collectionOf{\shpFuncPhiEnrichmentEps{I}}}
                {\collectionOf{\coeffPhiEnrichmentBar{I}}}, 
        \mathbf{R^\varepsilon}
    \bigg) d\bx^\varepsilon \\ 
=  \int_{\Omega} g^\varepsilon
        \bigg(
     \densOpMatBar,
        \bigg\{
            \wfnTildeDiscNumKEvalAt
                {\alpha}
                {\bk}
                {\collectionOf{\shpFuncClass{i}}}
                {\collectionOf{\coeffUClassicalBar{\alpha}{i}}}
                {\collectionOf{\shpFuncUEnrichmentEps{j}{I}}}
                {\collectionOf{\coeffUEnrichmentBar{\alpha}{j}{I}}}
        \bigg\},\\
        \phiTildeDiscEvalAt
                {\collectionOf{\shpFuncClass{j}}}
                {\collectionOf{\coeffPhiClassicalBar{j}}}
                {\collectionOf{\shpFuncPhiEnrichmentEps{I}}}
                {\collectionOf{\coeffPhiEnrichmentBar{I}}}, 
        \mathbf{R^\varepsilon}
    \bigg) d\bx \:,
\end{split} 
\end{equation}
where, $g^\varepsilon(\cdots) = f^\varepsilon(\cdots) det(\frac{\partial \bx^\varepsilon}{\partial \bx}).$ 

We note that the expression for the configurational force derived in Motamarri \& Gavini (2018) ~\cite{Motamarri2018} holds for classical finite element basis. However, it the case of enriched finite element basis, the enrichment functions in the perturbed space \big(\{($\shpFuncUEnrichmentEps{j}{I}$\}, \{$\shpFuncPhiEnrichmentEps{I}$\}\big) are dependent on $\varepsilon$. We perform the following simplification to arrive at the additional contributions arising from the enrichment functions: 

\begin{equation}
\begin{split}
\hat{F}^h(\boldsymbol{\Upsilon}(\bx)) =\int_{\Omega} 
\Bigg[
\sum_{\alpha}
\bigg\{
\parder{g^\varepsilon}{\wfnTildeDiscNumK{\alpha}{\bk}}\evalAtEpsZero 
\wfnSpecialEpsDer +
\parder{g^\varepsilon}{\wfnTildeDiscNumKConj{\alpha}{\bk}}\evalAtEpsZero
\wfnSpecialEpsDerConj +
\parder{g^\varepsilon}{\del\wfnTildeDiscNumKConj{\alpha}{\bk}}\evalAtEpsZero \cdot
\gradWfnSpecialEpsDerConj 
\\
+ \parder{g^\varepsilon}{\del\wfnTildeDiscNumK{\alpha}{\bk}}\evalAtEpsZero \cdot
\gradWfnSpecialEpsDer 
\bigg\} +
\parder{g^\varepsilon}{\phiTildeDisc}\evalAtEpsZero
\phiSpecialEpsDer +
\parder{g^\varepsilon}{\del\phiTildeDisc}\evalAtEpsZero \cdot
\gradPhiSpecialEpsDer 
\Bigg] d\bx  \\
+ \frac{d}{d\varepsilon} 
\int_{\Omega}
g^\varepsilon
        \bigg(
     \densOpMatBar,
        \bigg\{
            \wfnTildeDiscNumKEvalAt
                {\alpha}
                {\bk}
                {\collectionOf{\shpFuncClass{i}}}
                {\collectionOf{\coeffUClassicalBar{\alpha}{i}}}
                {\collectionOf{\shpFuncUEnrichment{j}{I}}}
                {\collectionOf{\coeffUEnrichmentBar{\alpha}{j}{I}}}
        \bigg\},\\
        \phiTildeDiscEvalAt
                {\collectionOf{\shpFuncClass{j}}}
                {\collectionOf{\coeffPhiClassicalBar{j}}}
                {\collectionOf{\shpFuncPhiEnrichment{I}}}
                {\collectionOf{\coeffPhiEnrichmentBar{I}}}, 
        \mathbf{R^\varepsilon}
    \bigg)
d\bx
\evalAtEpsZero\,.
\end{split}
\end{equation}
The last term in the above equation is denoted by $\hat{F}_{1}(\boldsymbol{\Upsilon}(\bx))$ in the main text (cf.~ Eq.~(\ref{eq_cfsplit})), which is the contribution to the configurational force from the classical finite element basis, and it is written out using the expressions presented in Motamarri \& Gavini (2018)~\cite{Motamarri2018}. The additional contribution arising from the enrichment functions is given by

\begin{equation}\label{eq_f2withg}
\begin{split}
\hat{F}_{2}(\boldsymbol{\Upsilon}(\bx)) =
\int_{\Omega} 
\Bigg[
\sum_{\alpha}
\bigg\{
\parder{g^\varepsilon}{\wfnTildeDiscNumK{\alpha}{\bk}}\evalAtEpsZero 
\wfnSpecialEpsDer +
\parder{g^\varepsilon}{\wfnTildeDiscNumKConj{\alpha}{\bk}}\evalAtEpsZero
\wfnSpecialEpsDerConj +
\parder{g^\varepsilon}{\del\wfnTildeDiscNumKConj{\alpha}{\bk}}\evalAtEpsZero \cdot
\gradWfnSpecialEpsDerConj 
+ \\
\parder{g^\varepsilon}{\del\wfnTildeDiscNumK{\alpha}{\bk}}\evalAtEpsZero \cdot
\gradWfnSpecialEpsDer 
\bigg\} +
\parder{g^\varepsilon}{\phiTildeDisc}\evalAtEpsZero
\phiSpecialEpsDer +
\parder{g^\varepsilon}{\del\phiTildeDisc}\evalAtEpsZero \cdot
\gradPhiSpecialEpsDer 
\Bigg] d\bx \:,
\end{split}
\end{equation}
where $\wfnSpecialEpsDer$ and $\phiSpecialEpsDer$ are defined in Eq.(\ref{eq_utilde_epsder}) and Eq.(\ref{eq_phitilde_epsder}), respectively. 

In the above equation, we also require the derivatives of $g^\varepsilon$ with respect to various fields, evaluated at $\varepsilon=0$, which we present next. For clarity, we drop the $\varepsilon$ suffix when we derive expressions for these derivatives.
We note that $g$ can be written as 
\begin{equation}
    g = g_{\text{KE}} + g_{\text{xc}} + g_{\text{elec}} + g_{\text{ent}} + g_{\text{const}} \:, 
\end{equation}
where $g_{\text{KE}}$, $g_{\text{xc}}$, $g_{\text{elec}}$, $g_{\text{ent}}$ and $g_{\text{const}}$ represent energy densities corresponding to the kinetic energy, exchange-correlation energy, electrostatic energy, entropic energy and the constraint, respectively.  
From Eq.~(\ref{eq_KineticEnergy}), $g_{\text{KE}}$ can be written as 
\begin{equation}
\begin{split}
    g_{\text{KE}}(\densOpMat,
                  \collectionOf{\wfnDiscNumK{\alpha}{\bk}},
                  \collectionOf{\del \wfnDiscNumK{\alpha}{\bk}}) \\
    = 2 \sum_{p, q, r=1}^{N} &
    \integrateBZ{
      \densOpMatElems{p}{q}
      \sMatElemInv{q}{r}
      \bigg(\frac{1}{2} \del \wfnDiscNumKConj{r}{\bk} \cdot \del \wfnDiscNumK{p}{\bk} 
          - i \wfnDiscNumKConj{r}{\bk} \bk.\del\wfnDiscNumK{p}{\bk}  
          + \frac{1}{2} |\bk|^2 \wfnDiscNumKConj{r}{\bk} \wfnDiscNumK{p}{\bk} \bigg)
    } \:.
\end{split}
\end{equation}
Hence,
\begin{equation}
\begin{split}
    \parder{g_{\text{KE}}}{\del \wfnDiscNumK{\alpha}{\bk}} 
    = 2  \sum_{p, q, r=1}^{N}
        \integrateBZ{
            \densOpMatElems{p}{q}
            \sMatElemInv{q}{r} 
            \bigg[ 
                \frac{1}{2} \del \wfnDiscNumKConj{r}{\bk}  \kronDelta{p}{\alpha}
                - i \wfnDiscNumKConj{r}{\bk} \bk \kronDelta{p}{\alpha}
            \bigg]
       } \:,
\end{split}
\end{equation}
where $\kronDelta{i}{j}$ is the Kronecker delta. When wavefunctions are given by the canonical (orthogonal) Kohn-Sham eigenfunctions, $\sMat$ is an identity matrix and $\densOpMat$ is simply a diagonal matrix containing fractional occupancies $\fracOcc{\alpha}{\bk}$. Hence, the above expression can be further simplified as follows:
\begin{equation}
   \parder{g_{\text{KE}}}{\del \wfnDiscNumK{\alpha}{\bk}} 
   = \integrateBZ{
       \fracOcc{\alpha}{\bk} \bigg[ 
            \del \wfnDiscNumKConj{\alpha}{\bk} - 2 i \wfnDiscNumKConj{\alpha}{\bk} \bk
            \bigg ]
   } = C_1 \:.
\end{equation}
On similar lines, we have 
\begin{equation}
   \parder{g_{\text{KE}}}{\del \wfnDiscNumKConj{\alpha}{\bk}} 
   = \integrateBZ{
       \fracOcc{\alpha}{\bk}  
            \del \wfnDiscNumK{\alpha}{\bk}   \:
   } = C_2 \:.
\end{equation}
The derivative of $g_{\text{KE}}$ with respect to the wavefunction is given by
\begin{equation}\label{eq_gKEwrtwfn}
\begin{split}
   \parder{g_{\text{KE}}}{\wfnDiscNumK{\alpha}{\bk}}  
   = 2  \sum_{p, q, r=1}^{N}
    \integrateBZ{
        \densOpMatElems{p}{q}
        \Bigg\{
        \funcDer{\sMatElemInv{q}{r}}{\wfnDiscNumK{\alpha}{\bk}}
        \bigg(
            \frac{1}{2} \del \wfnDiscNumKConj{r}{\bk} \cdot \del \wfnDiscNumK{p}{\bk} 
            - i \wfnDiscNumKConj{r}{\bk} \bk.\del\wfnDiscNumK{p}{\bk}  
            + \frac{1}{2} |\bk|^2 \wfnDiscNumKConj{r}{\bk} \wfnDiscNumK{p}{\bk} 
        \bigg) 
            + \\
        \sMatElemInv{q}{r}
        \frac{1}{2} |\bk|^2 \wfnDiscNumKConj{r}{\bk} \kronDelta{p}{\alpha}
        \Bigg\}
    }  \:.
\end{split}
\end{equation}
Using $\sMat \sMatInv = \matIdentity $, we can show the following in the context of orthogonal Kohn-Sham eigenfunctions:
\begin{equation}
\funcDer{\sMatInv}{\wfnDiscNumK{\alpha}{\bk}} = - \funcDer{\sMat}{\wfnDiscNumK{\alpha}{\bk}} \:.
\end{equation}
Using this result in Eq.(\ref{eq_gKEwrtwfn}), we have
\begin{equation}
\begin{split}
   \parder{g_{\text{KE}}}{\wfnDiscNumK{\alpha}{\bk}}  
   = A_1 + C_3 :\,
\end{split}
\end{equation}
where, 
\begin{equation}
   A_1 =  - 2  \sum_{p, q, r=1}^{N}
    \integrateBZ{
        \densOpMatElems{p}{q}
        \funcDer{\sMatElem{q}{r}}{\wfnDiscNumK{\alpha}{\bk}}
        \bigg(
            \frac{1}{2} \del \wfnDiscNumKConj{r}{\bk} \cdot \del \wfnDiscNumK{p}{\bk} 
            - i \wfnDiscNumKConj{r}{\bk} \bk.\del\wfnDiscNumK{p}{\bk}  
            + \frac{1}{2} |\bk|^2 \wfnDiscNumKConj{r}{\bk} \wfnDiscNumK{p}{\bk} 
        \bigg) 
    }
\end{equation}
and
\begin{equation}
    C_3 = 2 \integrateBZ{
        \fracOcc{\alpha}{\bk} 
        \frac{1}{2} |\bk|^2 \wfnDiscNumKConj{\alpha}{\bk} 
    } \:.
\end{equation}
Similarly, we can show that the derivative of $g_{\text{KE}}$ with respect to $\wfnDiscNumKConj{\alpha}{\bk}$ can be given by

\begin{equation}
\begin{split}
   \parder{g_{\text{KE}}}{\wfnDiscNumKConj{\alpha}{\bk}}  
   = B_1 + C_4 \:,
\end{split}
\end{equation}
where 
\begin{equation}
   B_1  = - 2  \sum_{p, q, r=1}^{N}
    \integrateBZ{
        \densOpMatElems{p}{q}
        \funcDer{\sMatElem{q}{r}}{\wfnDiscNumKConj{\alpha}{\bk}}
        \bigg(
            \frac{1}{2} \del \wfnDiscNumKConj{r}{\bk} \cdot \del \wfnDiscNumK{p}{\bk} 
            - i \wfnDiscNumKConj{r}{\bk} \bk.\del\wfnDiscNumK{p}{\bk}  
            + \frac{1}{2} |\bk|^2 \wfnDiscNumKConj{r}{\bk} \wfnDiscNumK{p}{\bk} 
        \bigg) 
    } \:,
\end{equation}
and
\begin{equation}
  C_4 =  2 \integrateBZ{
        \fracOcc{\alpha}{\bk} 
        \bigg( 
            -i \bk\cdot \del \wfnDiscNumK{\alpha}{\bk} + 
            \frac{1}{2} |\bk|^2 \wfnDiscNumK{\alpha}{\bk} 
        \bigg)
    } \:.
\end{equation}
We note that the derivative of $g_{\text{KE}}$ with respect to the potential $\phiDisc$ is zero. Before we move on to the terms beyond kinetic energy, we derive a useful result for the electron density.
Recall that the electron density (cf. Eq.(\ref{eq_ElectronDensity})) is given by
\begin{equation}
   \rhoDisc(\bx) 
   = 2 \sum_{p, q, r=1}^{N} \integrateBZ{
        \densOpMatElems{p}{q}
        \sMatElemInv{q}{r}
        \wfnDiscNumKConj{r}{\bk}
        \wfnDiscNumK{p}{\bk}
   }\:.
\end{equation}
Hence, we have the following results for the electron density:

\begin{equation}
\begin{split}
   \funcDer{\rhoDisc}{\wfnDiscNumK{\alpha}{\bk}} =  A_2 + C_5  
   \quad\quad \text{and} \quad\quad 
   \funcDer{\rhoDisc}{\wfnDiscNumKConj{\alpha}{\bk}} = B_2 + C_6 \:,
\end{split} 
\end{equation}
where,
\begin{equation}
    A_2 = - 2  \sum_{p, q, r=1}^{N}
    \integrateBZ{
        \densOpMatElems{p}{q}
        \funcDer{\sMatElem{q}{r}}{\wfnDiscNumK{\alpha}{\bk}}
        \wfnDiscNumKConj{r}{\bk} \wfnDiscNumK{p}{\bk} 
    } 
    \quad \quad \text{and} \quad \quad
    C_5 =  2 \integrateBZ{
        \fracOcc{\alpha}{\bk}
        \wfnDiscNumKConj{\alpha}{\bk}
    }
    \:, 
\end{equation}
and 
\begin{equation}
    B_2 = - 2  \sum_{p, q, r=1}^{N}
    \integrateBZ{
        \densOpMatElems{p}{q}
        \funcDer{\sMatElem{q}{r}}{\wfnDiscNumKConj{\alpha}{\bk}}
        \wfnDiscNumKConj{r}{\bk} \wfnDiscNumK{p}{\bk} 
    } 
    \quad \quad \text{and} \quad \quad
    C_6 = 2 \integrateBZ{
        \fracOcc{\alpha}{\bk}
        \wfnDiscNumK{\alpha}{\bk}
    } \:.
\end{equation}


Moving on to the other terms in Eq.~(\ref{eq_excorenergy}), beyond $g_{\text{KE}}$, we can write the energy density for the exchange-correlation energy as
\begin{equation} 
    g_{\text{xc}} =  F(\rhoDisc) \:.
\end{equation}
Therefore, the derivative with respect to the wavefunctions are given by
\begin{equation}
   \funcDer{g_{\text{xc}}}{\wfnDiscNumK{\alpha}{\bk}} = V_{\text{xc}} 
   \funcDer{\rhoDisc}{\wfnDiscNumK{\alpha}{\bk}} = C_7
   \quad\mathrm{and}\quad 
   \funcDer{g_{\text{xc}}}{\wfnDiscNumKConj{\alpha}{\bk}} = V_{\text{xc}} 
   \funcDer{\rhoDisc}{\wfnDiscNumKConj{\alpha}{\bk}} = C_8\:,
\end{equation}
where $V_{\text{xc}} = \funcDer{F}{\rhoDisc}$ is the exchange-correlation potential. The derivative of $g_{\text{xc}}$ with respect to the potential $\phiDisc$ is zero.
From Eq.~(\ref{eq_LocalElectrostaticsSmearedChargeFunctional}), the electrostatic energy density is given by
\begin{equation}
\begin{split}
     g_{\text{elec}}=    
        (\rhoDisc + b_s)\phiDisc - 
        \frac{1}{8\pi} \left|\del \phiDisc\right|^2 +    
        \sum_I  \rhoDisc ( V_{I} - V_{s,I} )   - 
        \sum_I \frac{1}{2}\,    b_{s,I}    V_{s,I}  \:.
\end{split}
\end{equation}
Hence, we can write derivatives of  $g_{\text{elec}}$ with respect to the wavefunctions as follows:
\begin{equation}
\begin{split}
\parder{g_{\text{elec}}}{\wfnDiscNumK{\alpha}{\bk}} 
    =  \funcDer{\rhoDisc}{\wfnDiscNumK{\alpha}{\bk}} 
        \bigg( 
            \phiDisc +   \sum_I  ( V_{I} - V_{s,I} )
        \bigg) = C_{9}
\quad \mathrm{and} \quad
\parder{g_{\text{elec}}}{\wfnDiscNumKConj{\alpha}{\bk}} 
    =  \funcDer{\rhoDisc}{\wfnDiscNumKConj{\alpha}{\bk}} 
        \bigg( 
            \phiDisc +   \sum_I  ( V_{I} - V_{s,I} )
        \bigg) = C_{10} \:.
\end{split}
\end{equation}
The derivative with respect to the electrostatic potential and its gradient are given by:
\begin{equation}
    \parder{g_{\text{elec}}}{\phiDisc}  
        =  \rhoDisc + b_s = C_{11}
    \quad \mathrm{and} \quad
    \funcDer{g_{\text{elec}}}{\del\phiDisc}  
        =  - \frac{1}{4\pi} \del \phiDisc = C_{12}\,.
\end{equation}
We note that $g_{\text{ent}}$ and  $g_{\text{const}}$ do not contribute to $\hat{F}_{2}(\boldsymbol{\Upsilon}(\bx))$ since they are not directly dependent on the wavefunctions or the electrostatic potential.

We now substitute all constituents derived above into Eq.(\ref{eq_f2withg}), and write the expression for $\hat{F}_{2}(\boldsymbol{\Upsilon}(\bx))$ as follows:
\begin{equation}\label{eqn:Appendix_finalSplit}
 \hat{F}_{2}(\boldsymbol{\Upsilon}(\bx)) = A + B + C \:.   
\end{equation}
In the above, $C$ includes all terms excepting those involving $A_1$, $A_2$, $B_1$ and $B_2$, and is given by
\begin{equation}\label{eqn_finalC}
\begin{split}
 C =    & \fint_{\Omega_{\text{BZ}}} \Bigg[  \sum_{\alpha} \bar{f}_{\alpha,\bk}  \int_{\Omega} \: \bigg\{ \del \, \Big( \left.  \frac{d}{d\varepsilon}  \widetilde{\ualbk^{h,\varepsilon \: *}}(\bx^\varepsilon)  \right \vert_{\varepsilon=0} \Big) \cdot \del \, \wfnDiscNumKBar{\alpha}{\bk}(\bx) + 
       \del \,  \wfnDiscNumKConjBar{\alpha}{\bk}(\bx)  \cdot \del \, \Big( \left. \frac{d}{d\varepsilon}\widetilde{\ualbk^{h,\varepsilon}}(\bx^\varepsilon) \right \vert_{\varepsilon=0} \Big)  \\
     & - 2i \, \epsder{\widetilde{\ualbk^{h,\varepsilon \: *}}(\bx^\varepsilon)} \bk\cdot\del \,  \wfnDiscNumKBar{\alpha}{\bk}(\bx) - 
       2i \, \wfnDiscNumKConjBar{\alpha}{\bk}(\bx) \bk\cdot\del \, \epsder{\widetilde{\ualbk^{h,\varepsilon}}(\bx^\varepsilon)} + |\bk|^2  \epsder{\widetilde{\ualbk^{h,\varepsilon \: *}}(\bx^\varepsilon)} \wfnDiscNumKBar{\alpha}{\bk}(\bx)  \\
     & \qquad \qquad + |\bk|^2   \wfnDiscNumKConjBar{\alpha}{\bk}(\bx) \epsder{\widetilde{\ualbk^{h,\varepsilon }}(\bx^\varepsilon)} \bigg\}   d\bx\Bigg] d\bk  \\
     & + \int_{\Omega} \big[ V_{\text{xc}}(\bx) + \hbarr{\phi}(\bx) + \sum_I  ( V_{I}(|\bx-\bR_I|) - V_{s,I}(|\bx-\bR_I|) )\big]
     \epsder{\widetilde{\rho^{h,\varepsilon}}(\bx^\varepsilon)} d\bx  \\
     &  + \intomega   (\hbarr{\rho}(\bx) + b_s(\bx,\bR)) \ddeps \widetilde{\phi^{h,\varepsilon}} (\eps{\bx}) \evalepszero d\bx  - \intomega \frac{1}{4\pi}     \del \big( \ddeps \widetilde{\phi^{h,\varepsilon}}(\eps{\bx})\big) \evalepszero \cdot \del \hbarr{\phi}(\bx)    d\bx \:,  
\end{split}
\end{equation}
where,
\begin{equation}
    \epsder{\widetilde{\rho^{h,\varepsilon}}(\bx^\varepsilon)} \coloneqq 2 \fint_{\Omega_{\text{BZ}}} \sum_{\alpha} \bar{f}_{\alpha,\bk} \Bigg[  
     \Big( \left.  \frac{d}{d\varepsilon}  \widetilde{\ualbk^{h,\varepsilon \: *}}(\bx^\varepsilon)  \right \vert_{\varepsilon=0} \Big)   \wfnDiscNumKBar{\alpha}{\bk}(\bx) + 
         \wfnDiscNumKConjBar{\alpha}{\bk}(\bx)    \Big( \left. \frac{d}{d\varepsilon}\widetilde{\ualbk^{h,\varepsilon}}(\bx^\varepsilon) \right \vert_{\varepsilon=0} \Big)
    \Bigg] d\bk \:.
\end{equation}
In Eq.~(\ref{eqn:Appendix_finalSplit}), $A$ and $B$ include all terms not accounted by $C$. These can be rearranged and written as follows:
\begin{equation}
\begin{split}
A + B = - 2  \sum_{p, q, r=1}^{N}
    \integrateBZ{
        \epsder{\widetilde{\sMatElem{q}{r}}}
        \densOpMatElems{p}{q}
        \intomega
        \Bigg(
            \frac{1}{2} \del \wfnDiscNumKConjBar{r}{\bk} \cdot \del \wfnDiscNumKBar{p}{\bk} 
            - i \wfnDiscNumKConjBar{r}{\bk} \bk.\del\wfnDiscNumKBar{p}{\bk}  
            + \frac{1}{2} |\bk|^2 \wfnDiscNumKConjBar{r}{\bk} \wfnDiscNumKBar{p}{\bk} \\
           +\bigg[ V_{\text{xc}}(\bx) + \hbarr{\phi}(\bx) + \sum_I  ( V_{I}(|\bx-\bR_I|) - V_{s,I}(|\bx-\bR_I|) \bigg] \wfnDiscNumKConjBar{r}{\bk} \wfnDiscNumKBar{p}{\bk}
        \Bigg) 
        d\bx
    }   \:,
\end{split}
\end{equation}
where the integral over $\Omega$ in the above equation reduces to the eigenvalues of the Kohn-Sham equation for canonical Kohn-Sham eigenfunctions $\{\wfnDiscNumKBar{\alpha}{\bk}\}$. Further, $\epsder{\widetilde{\sMatElem{}{}}}$ is a diagonal matrix with elements given by
\begin{equation}
\begin{split}
  \epsder{\widetilde{\sMatElem{\alpha}{\alpha}}} \coloneqq
     \intomega \Big( \left.  \frac{d}{d\varepsilon}  \widetilde{\ualbk^{h,\varepsilon \: *}}(\bx^\varepsilon)  \right \vert_{\varepsilon=0} \Big)   \wfnDiscNumKBar{\alpha}{\bk}(\bx) + 
         \wfnDiscNumKConjBar{\alpha}{\bk}(\bx)    \Big( \left. \frac{d}{d\varepsilon}\widetilde{\ualbk^{h,\varepsilon}}(\bx^\varepsilon) \right \vert_{\varepsilon=0} \Big)d\bx \:.
\end{split}
\end{equation}
Hence, we have 
\begin{equation}\label{eqn_final_APlusB}
A + B = -2\fint_{\Omega_{\text{BZ}}} \Bigg[  \sum_{\alpha} \bar{f}_{\alpha,\bk}  \int_{\Omega} \:    \epsalbkBar \bigg\{ \epsder{\widetilde{\ualbk^{h,\varepsilon \:*}}(\bx^\varepsilon)} \wfnDiscNumKBar{\alpha}{\bk}(\bx) + 
     \wfnDiscNumKConjBar{\alpha}{\bk}(\bx) \epsder{\widetilde{\ualbk^{h,\varepsilon}}(\bx^\varepsilon)} \bigg\}   d\bx\Bigg] d\bk \:.
\end{equation}

Using the results in Eq.~(\ref{eqn_finalC}) and Eq.~(\ref{eqn_final_APlusB}), we obtain the the expression for $\hat{F}_{2}(\boldsymbol{\Upsilon}(\bx))$ presented in Eq.~(\ref{eq_finalF2}).

\end{widetext}

\bibliography{ref}
\bibliographystyle{apsrev4-1}

\end{document}